\documentclass[12pt]{article}

\RequirePackage{amsmath,amssymb}

\usepackage{fullpage}

\usepackage{comment}
\usepackage{graphicx}
\usepackage{amsmath}
\usepackage{amsthm}
\usepackage{setspace}
\usepackage{enumerate}
\usepackage{subcaption}

\usepackage[mdyyyy]{datetime}
\usepackage[lowtilde]{url}
\usepackage[longnamesfirst]{natbib}

\usepackage[english]{babel}
\usepackage{graphics}
\usepackage{rotating}

\usepackage{times}
\usepackage{color}
\usepackage[allcolors=blue, colorlinks=true,linkcolor=blue, citecolor=blue]{hyperref}

\usepackage[table,xcdraw]{xcolor}

\usepackage{graphicx}
\usepackage{epstopdf}
\newcommand\numberthis{\addtocounter{equation}{1}\tag{\theequation}}

\theoremstyle{plain}	\newtheorem{theorem}{Theorem}[section]
\newtheorem{lemma}[theorem]{Lemma}

\theoremstyle{definition}

\DeclareSymbolFont{AMSb}{U}{msb}{m}{n}
\DeclareMathSymbol{\N}{\mathbin}{AMSb}{"4E}
\DeclareMathSymbol{\Z}{\mathbin}{AMSb}{"5A}
\DeclareMathSymbol{\R}{\mathbin}{AMSb}{"52}
\DeclareMathSymbol{\Q}{\mathbin}{AMSb}{"51}
\DeclareMathSymbol{\I}{\mathbin}{AMSb}{"49}
\DeclareMathSymbol{\C}{\mathbin}{AMSb}{"43}
\DeclareMathSymbol{\D}{\mathbin}{AMSb}{"44}
\DeclareMathSymbol{\E}{\mathbin}{AMSb}{"45}

\def\b1{\boldsymbol{1}}

\def\RR{\hbox{I\kern-.1667em\hbox{R}}}

\newcommand{\pr}[1]{\mathbf{P} \! \left\{ #1 \right\}}
\newcommand{\EE}[1]{\E \! \left [ #1 \right ]}
\newcommand{\EEsub}[2]{\E_{#1} \! \left [ #2 \right ]}

\def\Var{\mbox{Var}}
\newcommand{\var}[1]{\Var \! \left [ #1 \right ]}
\newcommand{\varsub}[2]{\Var_{#1} \! \left [ #2 \right ]}
\newcommand{\MSE}[1]{\mbox{MSE}\! \left [ #1 \right ]}

\newcommand{\cov}[1]{\mbox{Cov} \! \left [ #1 \right ]}

\newcommand*\rfrac[2]{{}^{#1}\!/_{#2}}

\usepackage{floatrow}
\newfloatcommand{capbtabbox}{table}[][\FBwidth]

\usepackage{blindtext}

\newif\ifissubmission

\issubmissionfalse

\begin{document}

\title{Worth Weighting? How to Think About and Use Weights in Survey Experiments}

\ifissubmission

\else

\author{Luke W. Miratrix\thanks{For research support and funding of the experimental studies analyzed here, Theodoridis thanks the University of California, Merced, the National Science Foundation Awards \#1430505, \#1225750 and \#0924191, the Empirical Implications of Theoretical Models program and the Integrative Graduate Education and Research Traineeship program, and the Berkeley Institute of Governmental Studies and its Mike Synar Graduate Research Fellowship; Miratrix thanks the Institute of Education Sciences, U.S. Department of Education, through Grant R305D150040; and Sekhon thanks Office of Naval Research (ONR) grants N00014-15-1-2367 and N00014-17-1-2176. The opinions expressed are those of the authors and do not represent views of the Institute, the U.S. Department of Education, ONR, or any of the organizations mentioned above. Please send comments to \href{mailto:lmiratrix@fas.harvard.edu}{\tt lmiratrix@stat.harvard.edu}, \href{mailto:sekhon@berkeley.edu}{\tt sekhon@berkeley.edu}, \href{mailto:atheodoridis@ucmerced.edu}{\tt atheodoridis@ucmerced.edu}, or \href{mailto:luiscampos@g.harvard.edu}{\tt luiscampos@g.harvard.edu}. Replication data happily provided upon request.} \\ Harvard University \and Jasjeet S. Sekhon \\ University of California, Berkeley \and Alexander G. Theodoridis \\ University of California, Merced \and Luis F. Campos \\ Harvard University} \date{} 

\fi 

\maketitle

\begin{abstract}

The popularity of online surveys has increased the prominence of using weights that capture units' probabilities of inclusion for claims of representativeness. 
Yet, much uncertainty remains regarding how these weights should be employed in analysis of survey experiments: Should they be used or ignored? If they are used, which estimators are preferred?  We offer practical advice, rooted in the Neyman-Rubin model, for researchers producing and working with survey experimental data. We examine simple, efficient estimators for analyzing these data, and give formulae for their biases and variances.  We provide simulations that examine these estimators as well as real examples from experiments administered online through YouGov.  We find that for examining the existence of population treatment effects using high-quality, broadly representative samples recruited by top online survey firms, sample quantities, which do not rely on weights, are often sufficient.  
We found that Sample Average Treatment Effect (SATE) estimates did not appear to differ substantially from their weighted counterparts, and they avoided the substantial loss of statistical power that accompanies weighting.  
When precise estimates of Population Average Treatment Effects (PATE) are essential, we analytically show post-stratifying on survey weights and/or covariates highly correlated with the outcome to be a conservative choice. While we show these substantial gains in simulations, we find limited evidence of them in practice.

\end{abstract}


\newpage

\doublespacing
\normalsize

\section{Introduction}

Population-based survey experiments have become increasingly common in political science in recent decades \citep{Gaines2007, Mutz2011, Sniderman2011}.  However, practical advice remains limited in the literature and uncertainty persists among scholars regarding the role of weights that capture differing probabilities of eventual inclusion across units in the analysis of survey experiments \citep{Franco2017}. Should they be used or ignored? If they are to be used, which estimators are to be preferred?  As \citet[][113-120]{Mutz2011} notes, 
\begin{quote}
	``there has been no systematic treatment of this topic to date, and some scholars have used weights while others have not ... the practice of weighting was developed as a survey research tool---that is, for use in observational settings. The use of experimental methodology with representative samples is not yet sufficiently common for the analogous issue to have been explored in the statistical literature.'' 
\end{quote}
We seek to fill this void with a systematic treatment, based on sound statistical principals rooted in the Neyman-Rubin model, yielding practical advice for scholars seeking to make the best possible decisions when using (or electing not to use) weights in their analysis of survey experiments.  We explore the topic through a combination of formulae, simulation, and examination of real data. 
 
 Taken together, these explorations lead to the conclusion that, for scholars examining population treatment effects using the high-quality, broadly representative samples recruited and delivered by top online survey firms, sample quantities, which do not rely on weights, are often sufficient.  Sample Average Treatment Effect (SATE) estimates tend not to differ substantially from weighted estimates, and they avoid the statistical power loss that accompanies weighting.  When precise estimates of Population Average Treatment Effects (PATE) are essential, we conclude that a ``double-H\`{a}jek'' weighted estimator is a very straightforward and reliable option in many cases.  We also analytically show that post-stratifying on survey weights and/or covariates highly correlated with the outcome is a conservative choice for precision improvement, because it is unlikely to do harm and could be quite beneficial in certain circumstances.

The greater prevalence of online surveys has gone hand-in-hand with the boom in survey experiments.  Firms such as YouGov (formerly Polimetrix) and Knowledge Networks (now owned by GfK) provide researchers platforms through which to run experiments. The firms offer representative samples generated through extensive panel recruitment efforts and sophisticated sample matching and weighting procedures. By reducing or eliminating costs, subsidized, grant-based and collective programs such as Time Sharing Experiments for the Social Sciences (TESS), the Cooperative Congressional Election Study (CCES), and Cooperative Campaign Analysis Project (CCAP) have further facilitated researchers' access to time on high-end online surveys.  
Other firms and platforms, such as Survey Sampling International, Google Consumer Surveys \citep{Santoso2016}, and Amazon's Mechanical Turk \citep{Berinsky2012}, offer even less costly access to large and diverse convenience samples on which researchers can also conduct survey experiments.  
Researchers using these sometimes generate their own weights to improve representativeness. 
However, because we view population inferences with such convenience samples as rather tenuous, our primary interest is in methods for analysis of data from sources, such as YouGov and Knowledge Networks, that actively recruit subjects and provide the researcher with weights.  

Survey experiments are a two-step process where a sample is first obtained from a parent population, and then that sample is randomized into different treatment arms. The sample selection and treatment assignment processes are generally independent of each other. Sampling procedures have changed in recent years because of increasing rates of non-response and new technologies. As a result, weights can vary substantially across units, with some units having only a small probability of being in the sample.  In contrast, the treatment assignment mechanisms are usually simple and relatively balanced, rendering the SATE straightforward to estimate. Estimating the PATE, however, is less so because these estimates need to incorporate the weights, which introduces additional variance as well as a host of complexities.     

In this work we assume the weights are known, and further assume that they incorporate both sampling probabilities and nonresponse. 
In particular, if there is non-response, and the non-response is correctly modeled as a function of some set of covariates, the overall weight would then be the product of being included in the sample and of responding conditional on that inclusion. 
We use \emph{weight} rather than \emph{sampling weight} to indicate this more general view.
In fact, for our primary type of data targeted by this work, typically the weights are calculated by the survey firms to represent the relative chances that a newly arrived recruit would get selected into the survey; as volunteering is in part self-selection, the non-response is built in to the final weight calculations automatically.
We believe the findings based on these assumptions are nevertheless informative, but we also discuss the additional complications of weight uncertainty in the body of this paper.

Overall, we encourage researchers choosing between these approaches to first give serious thought to the types of inferences they will make.  Do they simply wish to establish the presence or absence of an effect in a given population?  If so, the SATE may suffice.  Or do they hope to measure the magnitude of an effect that must or may not already be documented?  In this case, the scholar may consider her options for weighted estimators.



In Section \ref{sec:Overview_surveys} we overview general survey methodology. 
In Section \ref{sec:survery_experiments_and_SATE} we then formally consider survey experiments and relate them to the SATE. 
We formally define the PATE and some estimators of it in Section \ref{sec:estimating_PATE}, where we also discuss weights and uncertainty in weights in more detail, and we introduce a post-stratification estimator in Section \ref{sec:post-strat}. 
We then investigate the performance of these estimators through  simulation studies in Section \ref{sec:simulations}, and analyze trends and features of real survey experimental data collected through YouGov in Section \ref{sec:application}.  
We conclude with an extended discussion, providing some advice and high-level pointers to applied practitioners.

\section{Surveys and Survey Experiments through the Lens of Potential Outcomes}
\label{sec:Overview_surveys}

We formalize surveys and survey experiments in terms of the Neyman-Rubin model of potential outcomes \citep{SplawaNeyman:1990ux}. 
Assume we have a population of $N$ units indexed as $i = 1, \ldots N$.  We take a sample from this population using a sample selection mechanism, and we then randomly assign treatment in this sample using a treatment assignment mechanism. Both mechanisms will be formally defined in subsequent sections.  Each unit $i$ in the population has a pair of values, $(y_i(0), y_i(1))$, called its potential outcomes.  Let $y_i(1) \in \R$ be unit $i$'s outcome if it were treated, and $y_i(0)$ its outcome if it were not.  For each selected unit, we observe either $y_i(1)$ or $y_i(0)$ depending on whether we treat it or not.  For any unselected unit, we observe neither.

We make the usual no-interference assumption that implies that treatment assignment for any particular unit has no impact on the potential outcomes of any other unit. This assumption is natural in survey experiments. The treatment effect $\Delta_i$ for unit $i$ is then the difference in potential outcomes, $\Delta_i \equiv y_i(1) - y_i(0)$. These individual treatment effects are deterministic, pre-treatment quantities.

Let $\mathcal{S}$ be our sample of $n$ units.  Then the Sample Average Treatment Effect (SATE) is the mean treatment effect over the sample:
\begin{equation}
\label{eq:tau_S}
\tau_{\mathcal{S}} = \frac{1}{n} \sum_{i \in \mathcal{S}} \Delta_i = \frac{1}{n} \sum_{i \in \mathcal{S}} y_i(1) - \frac{1}{n} \sum_{i \in \mathcal{S}} y_i(0).
\end{equation}
This is a parameter for the sample at hand, but is random in its own right if we view the sample as a draw from the larger population.
By comparison, a parameter of interest in the population is the Population Average Treatment Effect (PATE) defined as
\[ \tau = \frac{1}{N} \sum_{i=1}^N \Delta_i = \frac{1}{N} \sum_{i=1}^N  y_i(1) - \frac{1}{N} \sum_{i=1}^N y_i(0)  . \]
In general $\tau_{\mathcal{S}} \neq \tau$, and if the sampling scheme is not simple (e.g., some types of units are more likely to be selected), then potentially $\EE{ \tau_{\mathcal{S}} } \neq \tau$.

We discuss some results concerning the sample selection mechanism in the next section. After that, we will combine the sample selection with the treatment assignment process.

\subsection{Simple Surveys (No Experiments)}
\label{subsec:survey_details}
Let $S_i$ be a dummy variable indicating selection of unit $i$ into the sample, with $S_i = 1$ if unit $i$ is in the sample, and 0 if not.  
Let $\mathcal{S}$ be $(S_1, \ldots, S_N)$, the vector of selections. 
In a slight abuse of notation, let $\mathcal{S}$ also denote the random sample.  
Thus, for example, $i \in \mathcal{S}$ would mean unit $i$ was selected into sample $\mathcal{S}$.
Finally let the overall \emph{selection probability} or \emph{sampling probability}  for unit $i$ be
\[ \pi_i \equiv \pr{ S_i = 1 } = \EE{ S_i },  \]
which is the probability of unit $i$ being included in the sample.  
The more the $\pi_i$ vary, the more the sample could be unrepresentative of the population.
We assume $\pi_i > 0$ for all $i$, meaning every unit has some chance of being selected into $\mathcal{S}$.
The $\pi_i$ depend, among other things, on the desired size of sample $\EE{ n }$.  
We assume the $\pi_i$ are fixed and known and incorporate non-response; we discuss uncertainty in them in Section~\ref{rev:weights}. \label{rev:pi_change} 
 
Consider the case where we have no treatment and we see $y_i \equiv y_i(0)$ for any selected unit.
Our task is to estimate the mean of the population, $\mu = \frac{1}{N} \sum_{i=1}^N y_i$.
Estimating the mean of a population under a sampling framework has a long, rich history.
We base our work on two estimators from that history here.  

Let $\bar{\pi} = \frac{1}{N} \sum \pi_i$ be the average selection probability in the population and $n = \sum S_i$ be the realized sample size for sample $\mathcal{S}$, with $\EE{ n } = N \bar{\pi}$. 
Then let $w_i = \bar{\pi}/\pi_i$ be the \emph{weight}.
These weights $w_i$ are relative to a baseline of 1, which eases interpretability due to removing dependence on $n$.
A weight of 1 means the unit stands for itself, a weight of 2 means the unit ``counts'' as 2 units, a weight of 0.5 means units of this type tend to be over-represented and so this unit counts as half, and so forth.
The total weight of our sample is then
\[  Z \equiv \sum_{i=1}^N \frac{\bar{\pi}}{\pi_i} S_i = \sum_{i=1}^N w_i S_i . \]
$Z$ is random, but $\EE{Z} = \EE{n}$.  

The Horvitz-Thompson estimator \citep{Horvitz:1952tp}, an inverse probability weighting estimator, is then
\[ \hat{y}_{HT} = \frac{1}{ \EE{ n } } \sum_{i=1}^N \frac{ \bar{\pi} }{ \pi_i } S_i y_i = \frac{1}{\EE{Z}}  \sum_{i=1}^N S_i w_i  y_i  . \]

Although unbiased, the Horvitz-Thompson estimator is well known to be highly variable.  This variability comes from the weights; if you randomly get too many rare units in the sample, the inverse of their weights will inflate $\hat{y}_{HT}$, even if all $y_i$ are the same.  We are not controlling for the realized size of the sample.  This is reparable by normalizing by the realized weight of the sample rather than the expected.

This gives the H\`{a}jek estimator, which is the usual weighted average of the selected units, and which likely reflects the approach used by most scholars:
\[ \hat{y}_{H} = \frac{1}{Z} \sum_{i=1}^N w_i S_i y_i \]

The H\`{a}jek estimator is not unbiased, but it often has smaller MSE than Horvitz-Thompson \citep{Hajek:1958, sarndal2003model}.
The bias, however, will tend to be negligible, as shown by the following lemma:

\begin{lemma}
\label{lemma:bias_sample}
[A variation on Result 6.34 of \cite{Cochran:1977}] Under a Poisson selection scheme, i.e. units sampled independently with individual probability $\pi_i$, the bias of the H\`{a}jek estimator is $O(1/\EE{n})$.  In particular, the bias can be approximated as 
\[	\EE{ \hat{y}_H } - \mu \  \dot{=}\  -\frac{1}{\E[n]} \Bigg( \frac{1}{N}\sum_{i = 1}^N (y_i - \mu) \frac{\bar{\pi}}{\pi_i} \Bigg) = -\frac{1}{\E[n]} \cov{ y_i, w_i }.
\]
\end{lemma}
See Appendix C for proof. 
The above shows that, for a fixed population, the bias decreases rapidly as sample size increases. 
If we sample with equal probability or if the outcomes are constant, the bias is 0. 
However, if the covariance between the weights and $y_i$ is large, the bias could potentially be large also.
In particular, the covariance will be large if rare units (those with small $\pi_i$) systematically tend to be outliers with large $y_i - \mu$ because, as weights are non-negative inverses of the $\pi_i$, their distribution can feature a long right tail that drives the covariance.

\section{Survey Experiments and SATE}
\label{sec:survery_experiments_and_SATE}

Survey experiments are surveys with an additional treatment assigned at random to all selected units.
Independent of $S_i$ let $T_i$ be a treatment assignment, with  $T_i = 1$ if unit $i$ is treated, 0 otherwise.  
The most natural such assignment mechanism for our context is Bernoulli assignment, where each responding unit $i$ is treated independently with probability $p$ for some $p$.
Another common mechanism is the classic complete randomization, when a $np$-sized simple random sample of the $n$ units is treated.
Regardless, we assume randomization is a separate process from selection.
In particular, we assume that randomization does not depend on the weights.

If our interest is in the SATE, then a natural estimator is Neyman's difference-in-means estimator of
\begin{eqnarray}
\label{eq:SATE}
 \hat{\tau}_{SATE} & = & \frac{1}{n_1} \sum_{i=1}^n T_i y_i  - \frac{1}{n-n_1} \sum_{i=1}^n (1-T_i) y_i ,
\end{eqnarray}
with $n_1$ the (possibly random) number of treated units \cite[see ][]{SplawaNeyman:1990ux}. 

This estimator is essentially unbiased for the SATE ($\EE{\hat{\tau}_{SATE}|\mathcal{S}} = \tau_{\mathcal{S}}$), but unfortunately, the SATE is not generally the same as the PATE and $\EE{ \tau_{\mathcal{S}} } \neq \tau$ in general.
The bias, for fixed $n$, is
\begin{align}
 bias( \hat{\tau}_{SATE} ) =  \EE{  \hat{\tau}_{SATE} }  - \tau &=  \frac{1}{N} \sum_{i=1}^N \left( \frac{\pi_i}{\bar{\pi}} - 1 \right) \Delta_i \nonumber  \\
  &= cov( \frac{\pi_i}{\bar{\pi}}, \Delta_i ) . \label{eq:bias_formula}
\end{align}
See Appendix C for derivation.
As units with higher $\pi_i$ will be more likely to be selected into $\mathcal{S}$, the estimator will be biased toward the treatment effect of these units.
Equation~\ref{eq:bias_formula} shows an important fact: if the treatment impacts are not correlated with the weights, then there will be no bias.
In particular, if the selection probabilities are all the same, or there is no treatment effect heterogeneity, then the bias of $\hat{\tau}_{SATE}$ for estimating the PATE will be 0.

The variance of $\hat{\tau}_{SATE}$, conditional on the sample $\mathcal{S}$, is well known, but we include it here as we use it extensively.
\begin{theorem}
\label{thm:var_sd}
Let sample $\mathcal{S}$ be randomly assigned to treatment and control with $\EE{T_i}=p$ for all $i$ with either a complete randomization or Bernoulli assignment mechanism.
The unadjusted simple-difference estimator $\hat{\tau}_{SATE}$ is unbiased\footnote{Nearly unbiased, that is.  Under randomizations where the estimator could be undefined (e.g., there is a chance of all units getting assigned to treatment, such as with Bernoulli assignment where $n_1$ is random and $\pr{n_1 = 0} > 0$ or $\pr{n_0 = 0} > 0$), this unbiasedness is conditional on the event of the estimator being defined. Because this probability is generally exponentially small the bias is as well, however.  See \cite{miratrix2013adjusting} for further discussion.} for the SATE, i.e. $\E[ \hat{\tau}_{SATE} | \mathcal{S}] = \tau_{\mathcal{S}}$. 
Its variance is
\begin{eqnarray}
 \var{\hat{\tau}_{SATE}|\mathcal{S}} & = & \frac{1}{n} \left[ ( \beta_1 + 1) \sigma_S^2(1) + (\beta_0+1) \sigma_S^2(0) + 2 \gamma_S \right] \label{eq:var_sd_simpleA} \\
 &=& \frac{1}{n} \left[  \beta_1 \sigma_S^2(1) + \beta_0 \sigma_S^2(0) - \sigma_S^2(\Delta) \right]
 \label{eq:var_sd_simple}
\end{eqnarray}
where $\sigma_S^2(z)$ and $\sigma_S^2(\Delta)$ are the variances of the individual potential outcomes and treatment effects for the sample, and $\beta_\ell = \EE{ n / n_\ell }$ are the expectations (across randomizations) of the inverses of the proportion of units in the two treatment arms.
\end{theorem}

If $n_1$ is fixed, such as with a completely randomized experiment, then $\beta_1 = 1/p$, $\beta_0 = 1/(1-p)$ and the above simplifies to Neyman's result of
\[  \var{\hat{\tau}_{SATE}|\mathcal{S}}  = \frac{1}{n} \left[  \frac{1}{p} \sigma_S^2(1) + \frac{1}{1-p} \sigma_S^2(0) - \sigma_S^2(\Delta) \right]
\]

\label{rev:R1_2}
For Bernoulli assignment, the $\beta_\ell$ are complicated because of the expectation of the random denominator, and there are mild technical issues because the estimator is undefined when, for example, $n_1 = 0$.  One approach is to use $n/\EE{ n_1 } = 1/p$ as an approximation for $\beta_1$.  This approximation is quite good; the bias is of high order for the same reasons as the bias for the H\`{a}jek estimator (see Lemma~\ref{lemma:bias_sample}).  Furthermore, the undefined issue is of small concern as the chance of $n_1 = 0$ is exponentially small; giving the estimator an arbitrary value (e.g., 0) if this rare event occurs, introduces only a small bias.  An alternate approach is to condition on the number of units treated: set $p = n_1/n$ and use Neyman's results.  Conditioning is a reasonable choice that we prefer.  It leads to a more accurate (and more readable) formula. For details, including formal definitions and the derivations, see \cite{miratrix2013adjusting}.

It is important to underscore that any SATE analysis on the sample, given a truly modeled treatment assignment mechanism, is valid.  I.e., such an analysis is estimating a true treatment effect parameter, the SATE.  
If $\EE{ \tau_{\mathcal{S}} } = \tau$ then any SATE analysis will be correct for PATE as well (although estimates of uncertainty may be too low if they do not account for variability in $\tau_{\mathcal{S}}$).
In particular, if there is a constant treatment effect, then $\tau_{\mathcal{S}} = \tau$ for any sample, and the SATE will be the PATE, and all uncertainty estimates for the SATE will be the same as for the PATE.\footnote{The estimated uncertainty will, however, depend on the sample $\mathcal{S}$.  For example, if $\mathcal{S}$ happens to have widely varying units, $ \hat{\tau}_{SATE}$ will have high variance and the sample-dependent SATE SE estimate should generally reflect that by being large to give correct coverage for $\tau_{\mathcal{S}}$.  Now, as this is true for any sample, the overall \emph{process} will have correct coverage.}
But a constant treatment effect is a large assumption.

\section{Estimating the PATE}
\label{sec:estimating_PATE}

Imagine we had both potential outcomes $y_i(1), y_i(0)$ for all the sampled $i \in \mathcal{S}$.
These would give us exact knowledge of the SATE, and we could also use this information, coupled with the weights, to estimate the PATE.
In particular, with knowledge of the $y_i(\ell)$ we have a sample of treatment effects:
\[ \Delta_i = y_i(1) - y_i(0) \mbox{ for } i \in \mathcal{S} . \]

We can use these $\Delta_i$ to estimate the PATE, $\tau$, with, for example, a H\`{a}jek estimator:
\begin{align}
\label{eq:nu_hajek}
 \nu_{\mathcal{S}} &= \frac{1}{Z} \sum S_i \frac{\bar{\pi}}{\pi_i} \Delta_i 
	 = \frac{1}{Z} \sum S_i \frac{\bar{\pi}}{\pi_i} y_i(1) - \frac{1}{Z} \sum S_i \frac{\bar{\pi}}{\pi_i} y_i(0).
\end{align}
This oracle estimator $\nu_{\mathcal{S}}$ is slightly biased, but the bias is small, giving $\EEsub{\mathcal{S}}{\nu_{\mathcal{S}}} \approx \tau$.
If we wanted an unbiased estimator, we could use a Horvitz-Thompson estimator by replacing $Z$ with $\EE{n} = N\bar{\pi}$, the expected sample size.

Unfortunately, we do not, for a given sample $\mathcal{S}$, observe $\nu_{\mathcal{S}}$. 
We can, however, estimate it given the randomization and partially observed potential outcomes.
Estimating the PATE is now implicitly a two-step process: estimate the sample dependent $\nu_{\mathcal{S}}$, which in turn estimates the population parameter $\tau$.
Under this view, we have two concerns.
First, we want to accurately estimate $\nu_{\mathcal{S}}$ using all the tools available to simple randomized experiments such as adjustment methods or, if we can control the randomization, blocking.
Second, we want to focus on a sample parameter $\nu_{\mathcal{S}}$ that is itself a good estimator of $\tau$.  
See Appendix A for a more formal treatment of this.

\subsection{Estimating \texorpdfstring{$\nu_{\mathcal{S}}$}{nu S}}
\label{sec:estimating_nuS}

Equation~\ref{eq:nu_hajek} shows that our estimator is the difference in weighted means of our treatment potential outcomes and our control potential outcomes.
This immediately motivates estimating these means with the units randomized to each arm of our study, as with the following ``double-H\`{a}jek'' estimator
\begin{align}
\label{eq:hh}
 \hat{\tau}_{hh} &=  \frac{1}{Z_1} \sum_{i=1}^N S_i T_i \frac{\bar{\pi}}{\pi_i}  y_i(1) -  \frac{1}{Z_0}  \sum_{i=1}^N S_i (1-T_i) \frac{\bar{\pi}}{\pi_i}  y_i(0)
\end{align}
with
\[ Z_1 = \sum_{i=1}^N S_i T_i \frac{\bar{\pi}}{\pi_i} \mbox{ and } Z_0 = \sum_{i=1}^N S_i (1-T_i) \frac{\bar{\pi}}{\pi_i} . \]
The $Z_\ell$ are the total sample masses in each treatment arm.  
$\EE{Z_1} = pN\bar{\pi} = \EE{n_1}$, the expected number of units that will land in treatment (similarly for control).

This estimator is two separate H\`{a}jek estimators, one for the mean treatment outcome and one for the mean control. 
Each estimator adjusts for the total mass selected into that condition. 
This difference of weighted means is the one naturally seen in the field. 
It corresponds to the weighted OLS estimate from regressing the observed outcomes $Y^{obs}$ on the treatment indicators $T$ with weights $w_i$. This equivalence is shown in Appendix C.

Because this is a H\`{a}jek estimator, there is bias for $\hat{\tau}_{hh}$ in the randomization step as well as the selection step because the $Z_\ell$ depend on the realized randomization.
Again, this bias is small, which means the expected value of our actual estimator, conditional on the sample, is approximately $\nu_{\mathcal{S}}$, our H\`{a}jek ``estimator'' of the population $\tau$: $\EE{ \hat{\tau}_{hh} | \mathcal{S} } \approx \nu_{\mathcal{S}}$.
(For unbiased versions, see Appendix A.)

We can obtain approximate results for the population variance of $\hat{\tau}_{hh}$ if we view the entire selection-and-assignment process as drawing two samples from a larger population.  
We ignore the finite-sample issues of no unit being able to appear in both treatment arms (i.e., we assume a large population) and use approximate formula based on sampling theory.  
For a Poisson selection scheme and Bernoulli assignment mechanism we then have:

\begin{theorem}
\label{thm:AV_hh}
The approximate variance (AV) of $\hat{\tau}_{hh}$ is	
\begin{align*}
 AV( \hat{\tau}_{hh} ) &\approx \frac{1}{p  \EE{n} } \frac{1}{N} \sum_{j=1}^N  w_j \left(y_j(1) - \mu(1) \right)^2 
     + \frac{1}{(1-p) \EE{n} }\frac{1}{N} \sum_{j=1}^N  w_j \left(y_j(0) - \mu(0) \right)^2,
\end{align*}
with $\mu(z) = \frac{1}{N}\sum_{i = 1}^N y_i(z)$.  This formula assumes the $\pi_j$ are small; see Appendix C for a more exact form.
This variance can be estimated by
\begin{align*}
\widehat{V}( \hat{\tau}_{sd} ) &= \frac{1}{Z_1^2}  \sum_{j = 1}^N S_i T_i w_j^2  \left(y_j(1) -  \hat{\mu}(1) \right)^2 
  + \frac{1}{Z_0^2}  \sum_{j = 1}^N S_i (1-T_i) w_j^2 \left(y_j(0) -  \hat{\mu}(0) \right)^2,
\end{align*}
where $\hat\mu(1) = \frac{1}{Z_1}\sum_{i = 1}^N S_i T_i w_i y_i(1)$ and $\hat\mu(0) = \frac{1}{Z_0}\sum_{i = 1}^N S_i (1-T_i) w_i y_i(0)$.
\end{theorem}
See Appendix C for the derivation, which also gives more general formulae that can be adapted for other selection mechanisms.
For related work and similar derivations, see \cite{aronow2013class} and \cite{wood2008covariance}.

\subsection{Uncertain and misspecified weights}
\label{rev:weights}

Following the survey sampling literature, this paper assumes the weights are exact, correct, and known.  They are considered to be the total chance of selection into the sample.  In particular, again following standard practice, the $\pi_i$ are the product of any original sampling weights and any non-response weights, given a classic sampling context \citep{groves2011survey}.  By contrast, for surveys such as YouGov the non-response is built in, as the recruited panels are in effect self-selected, so we get the overall weights (which they call propensity or case weights) directly. Our results are regarding these total weights.

Of course, especially when considering non-response, weights are not known but instead estimated using a model and, ideally, a rich set of covariates. 
This raises two concerns. 
The first is if the weights are systematically inaccurate due to some selection mechanism that has not been correctly captured. 
In this case, as the weights are independent of the assignment mechanism, the SATE estimates are still valid and unbiased. 
The PATE estimates, however, can be arbitrarily biased, and this bias is not necessarily detectable. 
For example, if only those susceptible to treatment join the study, the PATE estimate will be too high, and there may be no measured covariate that allows for detection of this.

The second concern is whether there is additional uncertainty that needs to be accounted for, given the estimated weights, when doing inference for the PATE.  There is, although we believe this uncertainty can often be much smaller than the uncertainty in the randomized experiment itself.\footnote{Consider that the estimated weights are usually calibrating the full sample to a larger known population, while the uncertainty of the experiment is of the difference in two subsamples, which will tend to have about four times the variance, at least.} While this uncertainty could be taken into account, much of the literature does not tend to do so.  Interestingly, it is not obvious whether estimating the weights given the sample could actually \emph{improve} PATE precision similar to using estimated propensity scores instead of known propensity scores---see, for example, \cite{Hirano:2000wo}.  We leave this as an area for future investigation.

For further thoughts on concerns regarding uncertainty in the weights, we point to the literature on generalizing randomized trials to wider populations, such as discussed in \cite{Hartman:2015ta} and \cite{Imai:2008vs}.
Here, the approach is generally to estimate units' propensity for inclusion into the experiment, and then weight units by these quantities in order to estimate population characteristics. 
These propensities of inclusion are usually estimated by borrowing from the propensity score literature for observational studies \citep{Cole:2010bf}.  
One nice aspect of this approach is it provides diagnostics in the form of a placebo test. 
In particular, the characteristics of the re-weighted control group of the randomized experiment should match the characteristics of the target population of interest (see \cite{Stuart:2010wm} for a discussion). 

Relatedly, \cite{OMuircheartaigh:2014wx} and \cite{Tipton:2012fj} propose post-stratified estimators, stratifying on these estimated weights.  
In their case, however, they also have the population proportions of the strata as given, which allows for simpler variance expressions and arguably less sensitivity to error in the weights themselves.  
Furthermore, they do not incorporate the unit-level weights once they stratify.  
\cite{Tipton:2012fj} investigates the associated bias-variance trade-offs due to stratification, and gives advice as to when stratification will be effective.

Generalization assumes we know the assignment mechanism, but not necessarily the sampling mechanism.  
There is some work on the reverse case, with estimated propensity scores of treatment and known weights, see \cite{DuGoff:2013be}.  
Here the final propensity weights are also treated as fixed for inference.

\subsection{Post-stratification to improve precision}

One can improve the precision of an experiment by adjusting for covariates. For an examination of this under the potential outcome framework, see, for example, \cite{Lin:2013ws}.
We use post-stratification for this adjustment.
In post-stratification, treatment effects are first estimated within each of a series of specified strata of the data and then averaged together with weights proportional to strata size \citep{miratrix2013adjusting}.

We use post-stratification because it relies on very weak modeling assumptions and naturally connects with the weighting involved in estimating the PATE.
See Appendix B for the overall framework and associated estimators.  
Other estimators that rely on regression and other forms of modeling are also possible, see \cite{zheng:2003} or, more recently, \cite{Si:2013dx}. 
For post-stratification, the more the mean potential outcomes vary between strata, the greater the gain in precision. 
And given that it is precisely when the weights and outcomes are correlated that we must worry about the weights, post-stratifying on them is a natural choice.
Such stratification is easy to implement: simply build $K$ strata pre-randomization (but not necessarily pre-sampling) by, e.g., taking the $K$ weighted quantiles of the $1/\pi_i$ as the strata.

When the units are divided into $K$ quantiles by survey weight, the cut-points of those quantiles depend on the realized weights of the sample.
Because this is still pre-randomization, this does not impact the validity of the variance and variance-estimation formulae of the SATE estimate of $\tau_{\mathcal{S}}$.
It does, however, make generating appropriate population variance formulae difficult.
Given this, we propose using the bootstrap, incorporating the variable definition of strata to take this stage being sample-dependent into account. 
Bootstrap is natural in that for survey experiments we are pulling units from a large population, and so simulating independent draws is reasonable.
While a technical analysis of this approach is beyond the scope of this paper, we discuss some particulars of implementation in Appendix B.
Reassuringly, our simulation studies in the next section show excellent coverage rates.

\section{Simulation Studies}
\label{sec:simulations}

We here present a series of simulation studies to assess the relative performance of the respective estimators.
We also assess the performance of the bootstrap estimates of the standard errors.

Our simulation studies are as follows: we generate a large population of size $N = 10,000$ with the two potential outcomes and a selection probability for each unit.  
Using this fixed population, we repeatedly take a sample and run a subsequent experiment, recording the treatment effect estimates for the different estimators.  
In particular, we first select a sample of size $n$, sampling without replacement but with probabilities of selection inversely proportional to the weights.\footnote{We ignore a mild technical issue of the $\pi_i$ not being exactly proportional to the weights due to not sampling with replacement.}
Once we have obtained the final sample, we randomly assign treatment and estimate the treatment effect.  
After doing this 10,000 times we estimate the overall mean, variance, and MSE of the different estimators to compare their performance to the PATE.
We also calculate bootstrap standard error estimates for all the estimators using the case-wise bootstrap scheme discussed in Appendix B.

\begin{table}[ht]
\centering
\begin{tabular}{rlrrrrrr}
 & Estimator & Mean & Bias & SE & RMSE & boot SE & Coverage \\ 
 A & \\
1 & $\tau_{\mathcal{S}}$ & 40.36 & 7.77 & 1.35 & 7.89  &  & \\ 
2 & $\nu_{\mathcal{S}}$ & 32.58 & 0.00 & 1.84 & 1.84 &  & \\ 
\hline
3 & $\hat{\tau}_{SATE}$ & 40.37 & 7.78 & 3.14 & 8.39 & 3.12 & 30\%
\\ 
4 & $\hat{\tau}_{hh}$ & 32.62 & 0.03 & 3.91 & 3.91 & 3.79  & 95\%\\ 
5 & $\hat{\tau}_{ps}$ & 32.60 & 0.01 & 2.67 & 2.67 & 2.69 & 95\%\\ 

 & \\
  
 B \\
1 & $\tau_{\mathcal{S}}$ & 30.00 & 0.00 & 0.00 & 0.00 &  & \\ 
2 & $\nu_{\mathcal{S}}$ & 30.00 & 0.00 & 0.00 & 0.00 &  & \\ 
\hline
3 & $\hat{\tau}_{SATE}$ & 30.01 & 0.01 & 2.58 & 2.58 & 2.58 & 95\%\\ 
4 & $\hat{\tau}_{hh}$ & 30.03 & 0.03 & 3.35 & 3.35 & 3.31 & 95\%\\ 
5 & $\hat{\tau}_{ps}$ & 30.02 & 0.02 & 3.32 & 3.32 & 3.29 & 95\%\\

\end{tabular}
\caption{Simulations A \& B. Performance of different estimators as estimators for the PATE for (A) a heterogenous treatment effect scenario with $\tau = 32.58$ and (B) a constant treatment effect of $\tau = 30$. For each estimator, we have, from left to right, its expected value, bias, standard error, root mean squared error, average bootstrap SE estimate and coverage across 10,000 trials.} 
\label{table:simulation_A}
\end{table}

\paragraph{Simulation A.}
Our first simulation is for a population with a heterogeneous treatment effect that varies in connection to the weight.
See Appendix D for some simple plots showing the structure of the population and a single sample. 
Our treatment effect, outcomes and sampling probabilities are all strongly related.
We then took samples of a specified size from this fixed population, and examined the performance of our estimators as estimators for the PATE.

Results for $n=500$ are on Table~\ref{table:simulation_A}. 
Other sample sizes such as $n=100$, not shown, are substantively the same.
The first two lines of the table show the performance of the two ``oracle'' estimators $\tau_{\mathcal{S}}$ (Equation \ref{eq:tau_S}) and $\nu_{\mathcal{S}}$ (Equation \ref{eq:nu_hajek}), which we could use if all of the potential outcomes were known. 
For $\tau_{\mathcal{S}}$ there is bias because the treatment effect of a sample is not generally the same as the treatment effect of the population. 
The H\`{a}jek approach of $\nu_{\mathcal{S}}$, second line, is therefore superior despite the larger SE.
 Line 3 is the simple estimate of the SATE from Equation \ref{eq:SATE}. Because it is estimating $\tau_{\mathcal{S}}$, it has the same bias as line 1, but because it only uses observed outcomes, the SE is larger. 
Line 4 uses the ``double-H\`{a}jek'' estimator shown in Equation~\ref{eq:hh}. 
This estimator is targeting $\nu_{\mathcal{S}}$, reducing bias, but has a larger SE relative to line 3 due to the fact that we are incorporating weights. 
Line 5 is the post-stratified ``double-H\`{a}jek'' given in Appendix B. 
Units were stratified by their survey weight, with $K = 7$ equally sized (by weight) strata. 
For this scenario, post-stratifying helps, as illustrated by the smaller SE and RMSE, compared to $\hat{\tau}_{hh}$.

An inspection of the coverage rates reflects what we have already discussed: The estimate $\hat\tau_{SATE}$ does not target the PATE while the other two sample estimates, $\hat\tau_{hh}$ and $\hat\tau_{ps}$, do. 
Therefore it has terrible coverage.
Furthermore, the latter two estimates give correct coverage, which is reflective of the bootstrap SE estimates hitting their mark.

\paragraph{Simulation B.}
As a second simulation we kept the original structure between $Y(0)$ and $w$, but set a constant treatment effect of 30 for all units.
Results are on the bottom half of Table~\ref{table:simulation_A}. 
Here, $\tau_{\mathcal{S}} = \tau$ for any sample $\mathcal{S}$, so there is no error in either estimate with known potential outcomes (lines 1--2).
This also means that $\hat{\tau}_{SATE}$ is a valid estimate of the PATE and this is reflected in the lack of bias and nominal coverage rate (line 3). The increase of SE of the weighted and post-stratified estimators (lines 4--5) reflects the use of weights when they are in fact unnecessary. 
Overall the SATE estimate is the best, as expected in this situation.

\begin{figure}[ht!]
\centering
	\includegraphics[width = 1\textwidth]{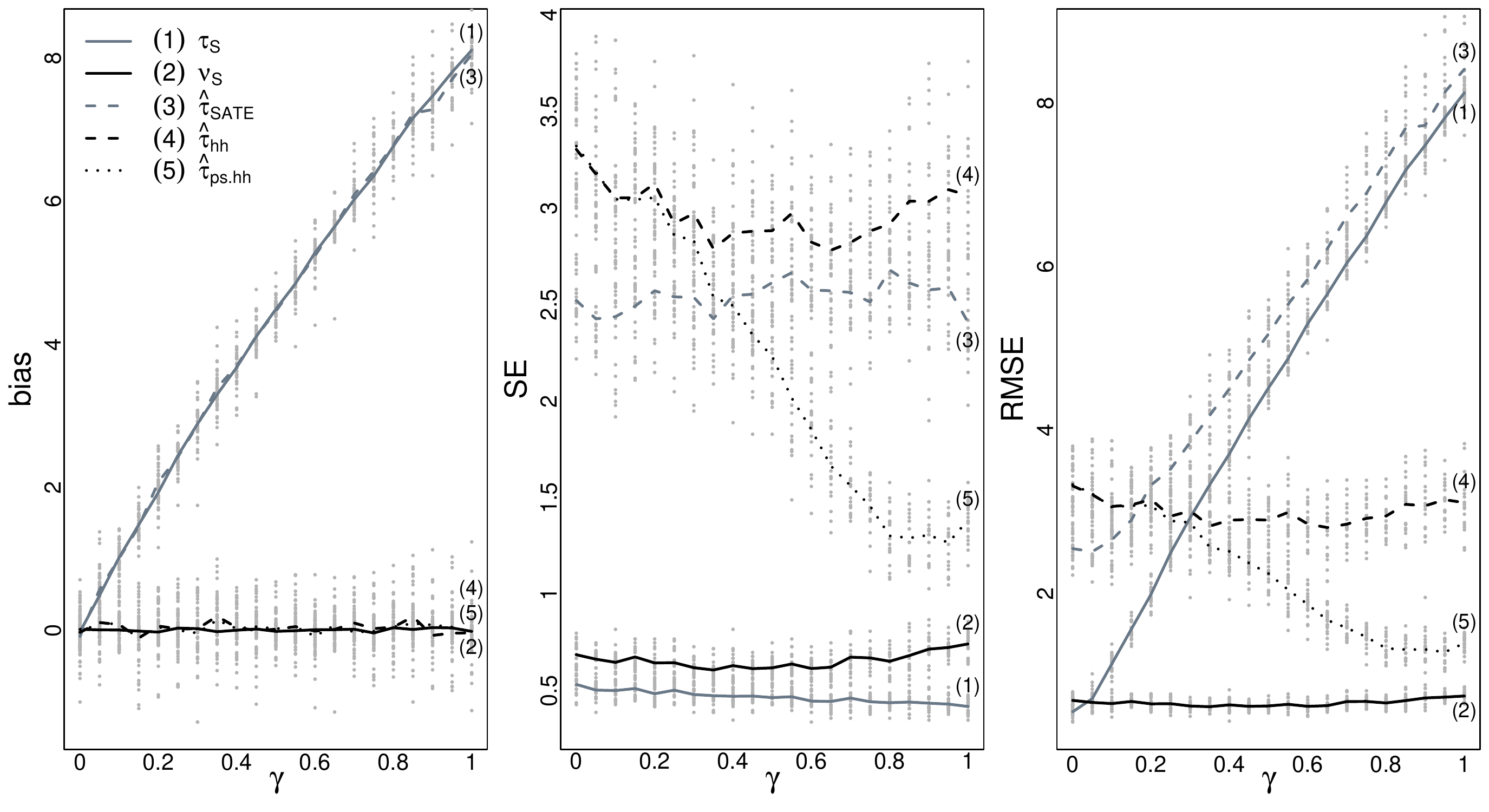}
\caption{Simulation C: (left to right) SE, bias and RMSE of estimates when selection probability is increasingly related to the potential outcomes. Grey are SATE-targeting, black PATE-targeting.  Solid are oracle estimators using all potential outcomes of the sample, dashed are actual estimators.  The thicker lines are averages over the 20 simulated populations in light gray dots.}
\label{fig:sim_D}
\end{figure}

\paragraph{Simulation C.}
In our final simulation, we systematically varied the relationship between selection probability and outcomes while maintaining the same  marginal distributions in order to examine the benefits post-stratification. 

In our DGP we first generate a bivariate normal pair of latent variables with correlation $\gamma$, and then generate the weights as a function of the first variable and the outcomes as a function of the second.
Then, by varying $\gamma$ we can vary the strength of the relationship between outcome and weight.
(See Appendix D for the particulars.) 
When $\gamma = 1$, which corresponds to Simulation A, $w_i$ and $(Y_i(0), Y_i(1))$ have a very strong relationship and we benefit greatly from post-stratification. 
Conversely when $\gamma = 0$, $w_i$ and $(Y_i(0), Y_i(1))$ are unrelated and there will be no such benefit. 

For each $\gamma$ we generated 20 populations, conducting a simulation study within each population.
We then averaged the results and plotted the averages against $\gamma$ on Figure~\ref{fig:sim_D}.
The solid lines give the performance of the oracle estimators $\tau_{\mathcal{S}}$ and $\nu_{\mathcal{S}}$, and the non-solid lines are the estimators. 
The gray lines are estimators that do not incorporate the weights, and the black lines are estimators that do.
The light grey points show the individual population simulation studies; they vary due to the variation in the finite populations.

We first see that, because both the double-H\`{a}jek and its post-stratified version are targeting $\nu_{\mathcal{S}}$, which in turn estimates the PATE, they remain unbiased regardless of the latent correlation. 
On the other hand, the SATE and its estimator, $\hat{\tau}_{SATE}$, are affected. 
The bias continually increases as the relationship between weight and treatment effect increases. 

As expected, the SE of the estimators that do not use weights, ${\tau}_{\mathcal{S}}$ and $\hat{\tau}_{SATE}$, stay the same regardless of $\gamma$ because the marginal distributions of the outcomes are the same across $\gamma$.
The estimators that only use weights to adjust for sampling differences, ${\nu}_{\mathcal{S}}$ and $\hat{\tau}_{hh}$, also remain the same, although their SEs are larger than for $\tau_{\mathcal{S}}$ and $\hat{\tau}_{SATE}$ because of incorporating the weights. 
We pay for unbiasedness with greater variability.
The post-stratified estimator $\hat{\tau}_{ps}$, however, sees continual precision gains as the weights are increasingly predictive of treatment effect.
For low $\gamma$, it has roughly the same uncertainty as $\hat{\tau}_{hh}$, but is soon the most precise of all (non-oracle) estimators. 

These conclusions are tied together in the right-most panel of Figure \ref{fig:sim_D}, showing the RMSE, which gives the combined impact of bias and variance on performance.
As $\gamma$ increases, the RMSE of $\hat{\tau}_{SATE}$ steadily climbs due to bias, eventually being the worst at $\gamma = 0.2$.
Meanwhile, the post-stratified estimator that exploits weights, $\hat{\tau}_{ps}$, performs better and better.
Overall, if weights are important then 1) the bias terms can be too large to be ignored, and 2) there is something to be gained by adjusting the estimates of treatment effects with those weights beyond simple re-weighting.
Otherwise, SATE estimators are superior, as incorporating weights can be costly.



\section{Real Data Application}
\label{sec:application}

To better understand the overall trade offs involved in using weighted estimators of PATE versus simply estimating the SATE on actual survey experiments, we analyzed a set of survey experiments embedded in 7 separate surveys fielded by us though YouGov over the course of 5 years. Studies appeared in two modules of the 2010 Cooperative Congressional Election Study (CCES), one module each of the 2012 and 2014 CCES, a survey of Virginia voters run prior to the 2013 gubernatorial election in that state, and two other national YouGov surveys. Each survey had post-hoc weights assigned by YouGov through that firm's standard procedure.  Across these surveys were 18 separate assignments of respondents to binary treatments.  In several of these cases, multiple outcomes were measured, producing 46 randomization/outcome combinations.\footnote{In the analyses presented here, we use all of these outcomes.  We elected to do this because more outcomes provide more opportunity for divergence between SATE and PATE, and thus a more conservative test of our conclusion that such divergence is rare.  Also, selecting outcomes would represent an added researcher degree of freedom, which we sought to avoid.  In the interest of transparency, we present, in Appendix E, the results of our examination when only the primary outcome for each randomization is used.  Our findings remain the same.}  All of the studies examined were conducted in the United States and focused on topics related to partisan political behavior.\footnote{Some of the studies featured random assignment of campaign advertisements shown to respondents with ad tone and partisan source varied.  Several of the studies presented respondents with vignettes or news stories describing candidates or groups of voters with characteristics such as party label and gender randomized.  Another set of studies asked respondents to evaluate artwork when told (or not told) that the art was produced by Presidents Bush and Obama.  More details regarding the specific studies used can be found in Table \ref{studydetails} in the Appendix.}  As such, we make our comparisons of SATE and PATE by looking at Democratic and Republican respondents separately.  This is because treatment effects for such studies are generally highly heterogeneous by respondent party identification.  
Our set of 46 randomization/outcome combinations produce 92 experiments (half among Democrats and half among Republicans).  
Sample sizes for the experiments analyzed range from 145 to 504 .
Weights varied substantially in these samples, ranging from near 0 to around 8, when normalized to 1 across the sample (standard deviation of 1.04).
Sixty-five of them ($70.7\%$) showed SATEs that were significantly different from zero.  
However, once the weights provided by YouGov were taken into account to estimate the PATE (via the double-H\`{a}jek estimate) only 52 experiments ($56.5\%$) had significant effects.  

\begin{figure}
	\centering
 	\includegraphics[width = 0.8\textwidth]{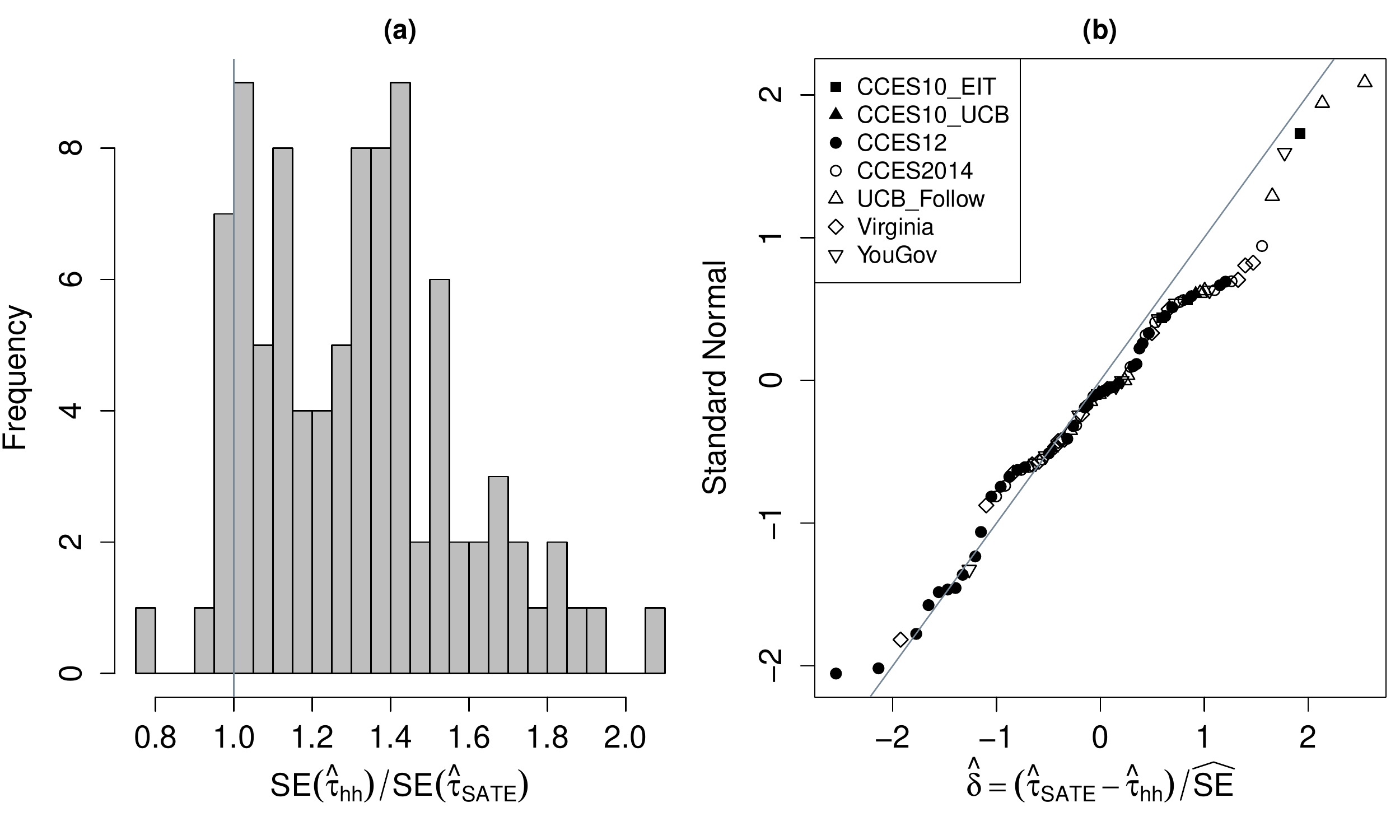}
	\caption{(a) Standardized efficiency of $\hat\tau_{hh}$ vs $\hat\tau_{SATE}$ (b) quantile-quantile plot of the {\emph{relative differences}} $\hat{\delta}$ of the estimates for the 92 experiments grouped into the larger surveys they are part of.}
	\label{fig:application_plot}
 \end{figure}

Our first finding is that incorporating weights substantially increased the standard errors.  
Figure \ref{fig:application_plot}(a) shows a $32.1\%$ average increase in standard error estimates of $\hat\tau_{hh}$ over $\hat\tau_{SATE}$ across experiments.

We next examined whether there is evidence of some experiments having a PATE substantially different from the SATE.
To do this, we calculated bootstrap estimates of the standard error for the difference in the estimators, and calculated a standardized difference in estimates of $\hat\delta = (\hat{\tau}_{SATE} - \hat{\tau}_{hh})/\widehat{SE}$.
If there were no difference between the SATE and the PATE, the $\hat\delta$s should be roughly distributed as a standard Normal.
First, the average of these $\hat{\delta}$ is $-0.115$, giving no evidence for any systematic trend of the PATE being larger or smaller than the SATE.
Second, when we compared our 92 $\hat\delta$ values to the standard normal with a qq-plot (Figure \ref{fig:application_plot}(b)), we find excellent fit.
While there is a somewhat suggestive tail departing from the expected line, the bulk of the experiments closely follow the standard normal distribution, suggesting that the SATE and PATE were generally quite similar relative to their estimation uncertainty.
A test using q-statistics (modeled after \cite{Weiss:2017tp}) failed to reject the null of no differences across the experiments ($p > 0.99$); especially considering the possible correlation of outcomes would make this test anti-conservative, we have no evidence that the PATE and SATE estimates differ (see Appendix E for further investigation of this).
An FDR test also showed no experiments with a significant difference.

Finally, we consider whether post-stratification on weights improved precision.  Generally, it did not: the estimated SEs of $\hat\tau_{ps}$ are very similar to those for $\hat\tau_{hh}$, with an average increase of about $0.6\%$.  
Further examination offers a hint as to why post-stratification did not yield benefits: the weights generated by YouGov for these samples do not correlate meaningfully with the outcomes of interest.  In no case did the magnitude of the correlation between weights and outcome exceed $0.23$.  
To further explore potential benefits of post-stratification we examine the effects of adjusting for a covariate, respondent party identification, on the full experiments not separated by party ID. 
Relative to no stratification, if we post-stratify on party ID, weighting the strata by the total sampled mass in both treatment and control, our PATE estimate shows an average reduction of $1.4\%$ in estimated SEs across experiments with participants of both major parties.
If we post-stratify on both party ID and the weights, we see an average estimated SEs reduction of $1\%$. These reductions would in no way make up for the larger SEs from attempting to estimate the PATE. 
However, they are consistent with the fact that post-stratification can generally only help.  

While this does not imply that scholars need not consider post-stratifying on weights, it does show that outcomes of interest in political science studies are not necessarily going to be correlated with these weights.  
This makes clear the importance of researchers understanding, and reporting, the process used to generate weights and being aware of the covariates with which those weights are likely to be highly correlated (for online surveys, such a list would often includes certain racial and education-level categories).  


\paragraph{Discussion.}
Overall, it appears that in this context and for these experiments,
the survey weights significantly increase uncertainty, and that there
is little evidence that the RMSE (which includes the SATE-PATE bias)
for estimating the PATE is improved by estimators that include these
weights.  Furthermore, the weights are not predictive enough of
outcome to help the post-stratified estimator.  With regard to
post-stratification, we note that in practice the analysis of any
particular experiment would likely be improved by post-stratifying on
known covariates predictive of outcome rather than na\"ively on the weights.

In understanding these findings, it is useful to consider the ways in which data from these leading online survey firms (in this case, YouGov) may differ from more convenience-based online samples.  Even unweighted, datasets from these firms tend to be more representative.  This is because they often engage in extensive panel recruitment and retention efforts and assign subjects from their panels to client samples through mechanisms such as block randomization.  As a result, the unweighted data are often largely representative of the overall population along many relevant dimensions.  Relatedly, firms may use a clean-up matching step, such as the one employed by YouGov, where they down-sample their data to generate more uniform weights  \citep{Rivers2006}.  This will likely increase the heterogeneity of the final sample, which could decrease precision.\footnote{Consider a standard scenario wherein a researcher purchases a sample of 1000 respondents.  To generate these data, the survey firm might recruit 1400 respondents, all of whom participate in the study. Two datasets result from this. The first contains all 1400 respondents. The second is a trimmed version, where the firm drops 400 of the most overrepresented respondents (which is tantamount to assigning these respondents a weight of 0).  This second set, which comes with weights assigned to each observation, is what many scholars analyze. Some firms will, upon request, also provide the full data set, but these data do not generally include weights, as the process for generating these weights is combined with the procedure for trimming down the larger data set by matching it to some frame based upon population characteristics. The weights will be less extreme than they would have been had the entire original sample been included, and the trimmed sample will be more heterogeneous, as many similar observations will be purged. This will make it more difficult to estimate its SATE compared to the full set (do note the SATEs could differ). Furthermore, post-stratification shows that estimators that include weights for the trimmed set will also be less variable than for the same estimators on the full dataset (assuming weights could be obtained), even though the trimmed dataset weights will be less variable.  Consider a case with two classes of respondents, reluctant and eager, equally represented in the population. The trimmed sample will have fewer eager respondents. Then, compared to the full data set, we will have a less precise estimate of the eager respondents in the trimmed data set.  The precision for the reluctant respondents would be the same.  Overall, our combined estimate will be, therefore, less precise.}   
We recommend that researchers request the original, pre-weighted data, in order to work with a larger and more homogeneous sample.  For the SATE the gains are immediate. For the PATE, one might generate weights for the full sample by extrapolating from the weights assigned in the trimmed sample or by contracting with the survey firm to obtain weights for this full unmatched sample. Then, by post-stratifying on the weights, the researcher can take advantage of the additional units to increase precision in some strata without increasing variability in the others.  For both SATE and PATE estimation, power would be  improved.


\section{Discussion and Practical Guidance}
\label{sec:conclusions}

We investigate incorporating weights in survey experiments under the potential outcomes framework.  We focus on two styles of estimator, those that incorporate these weights to take any selection mechanisms into account, and those that ignore weights and instead focus on estimating the SATE.  We primarily find that incorporating weights, even when they are exactly known, substantially decreases precision.  Because of this, researchers are faced with a trade-off: more powerful estimates for the SATE, or more uncertain estimates of the PATE.
We conclude with several observations that should inform how one navigates this trade-off.

The PATE can only be different from the SATE when two things hold: (1) there is meaningful variation in the treatment impact, and (2) that variation is correlated with the weights.  See Equation~\ref{eq:bias_formula}.  Moreover, the random assignment of treatment protects inference for the weighted estimator, even if the weights are incorrect or known only approximately: because the randomization of units into treatment is independent of the (possibly incorrect) weights, any inference conditional on the sample and the weights is a valid inference. When PATE is the estimand, we are estimating the treatment effect for a hypothetical population defined by the weights and sample, even if it does not correspond to the actual population.  For example, if we find a treatment effect in our weighted sample, we know the treatment does have an effect for at least some units. See \cite{Hartman:2015ta} for a discussion of this issue in the case of evaluating the external validity of an experiment.

It is important to compare the PATE and SATE estimates.  A meaningful discrepancy between them is a signal to look for treatment effect heterogeneity and a flag that weight misspecification could be a real concern.  If the estimates do not differ, however, and there is no other evidence of heterogeneity, then extrapolation is less of a concern---and furthermore the SATE is probably a sufficient estimate for the PATE.  Of course, with misspecified weights if there is heterogeneity associated with being selected into the experiment, but that is not captured by the covariates, then PATE estimation can be undetectably biased. For more on assessing heterogeneity, see \cite{ding2016decomposing}.


Interestingly, our examination of real survey data found no strong connection between the weights and outcomes. 
The SATE and PATE estimators tended to be similar.
Based on this, we have several general pieces of practical guidance: 
(1) When analyzing survey experiments using high quality, broadly representative samples, such as those recruited and provided by firms like YouGov and Knowledge Networks, SATE estimates will generally be sufficient for most purposes.  
(2) If a particular research question calls for estimates of the PATE, a ``double-H\`{a}jek'' estimator is probably the most straightforward (and a defensible) approach, unless weights are highly correlated with the outcomes variables.  (3) If weights are strongly correlated with a study's outcome(s) of interest, post-stratification on the weights with bootstrap standard errors can help offset the cost of including weights for those seeking to draw population inferences.  

This motivates a two-stage approach: first focus on the SATE using the entire, unweighted sample and determine whether the treatment had impact.  This will generally be the most powerful strategy for detecting an effect, as the weights, being set aside, will not inflate uncertainty estimates.  Then, once a treatment effect is established, work on how to generalize it to the population.  This second stage is an assessment of the magnitude of an effect in the population once an effect on at least some members of the population has been established.  First estimate the PATE with the weights, and then compare it to the SATE estimate.  If they differ, then consider working to explain any treatment effect heterogeneity with covariates, and think carefully about weight quality.  Regardless, ensure that all analyses preserve the original strength of the assignment mechanism; the weights do not need to jeopardize valid assessment of the presence of causal effects. Part of preserving valid statistical inference would be to commit to a particular procedure before analyzing the given dataset. A pre-analysis plan or sample splitting would help prevent a fishing expedition to find treatment effects.

%

\ifissubmission

\else

\section*{Acknowledgements}
For helpful and careful comments we would like to thank Henry Brady, Devin Caughey, Christopher Fariss, Erin Hartman, Steve Nicholson, Liz Stuart, Chelsea Zhang, and participants of the ACIC 2015 and 2014 Society of Political Methodology Summer Meetings. We would like to thank Guillaume Basse for his insights into the connection between weighted linear regression and the double-H\`{a}jek estimator. We also thank the valuable feedback and commentary received from two anonymous reviewers and the Editor, who pushed us to clarify and refine our findings.

\fi

\newpage

\bibliographystyle{apsr}
\bibliography{references}

\newpage

\section*{Appendix A: A general class of estimators}

When estimating the PATE, our overall estimation error is a combination of our error due to the randomized experiment for estimating $\nu_{\mathcal{S}}$ and the difference between our survey-sampling estimate $\nu_{\mathcal{S}}$ and the PATE $\tau$.
We can break this error down for any estimator $\hat{\tau}_*$ of $\nu_{\mathcal{S}}$.  
First, given $\hat{\tau}_*$, we have $\EE{ \hat{\tau}_* | \mathcal{S} } = \nu_{\mathcal{S}} + b_S$, with $b_S$ being a bias term.
Then
\begin{align*}
 \MSE{ \hat{\tau}_* } &= \EE{ (\hat{\tau}_* - \tau)^2 } \\
		&= \EE{ (\hat{\tau}_* - \nu_{\mathcal{S}})^2 } + \EE{ (\nu_{\mathcal{S}} - \tau)^2 } + 2\EEsub{\mathcal{S}}{ b_S(\nu_{\mathcal{S}} - \tau)} \numberthis \label{eq:mse_general} \\
		&= \EEsub{\mathcal{S}}{ \MSE{\hat{\tau}_* | \mathcal{S}}} + \MSE{ \nu_{\mathcal{S}} } + 2\EEsub{\mathcal{S}}{ b_S(\nu_{\mathcal{S}} - \tau)} 
\end{align*}
Given a choice of $\nu_{\mathcal{S}}$, the first term is the expected MSE of the estimator for estimating $\nu_{\mathcal{S}}$ when we consider all possible \emph{randomizations} of treatment assignment on the given sample $\mathcal{S}$.   
The second term is the MSE of $\nu_{\mathcal{S}}$ as an estimator for $\tau$ across all samples.
The third term is a cross-bias term; it depends on how the bias of a sample is correlated with the error of its $\nu_{\mathcal{S}}$.
We generally assume it is small and ignore it.  This gives a rough formula for the overall mean square error of
\begin{align}
 \MSE{ \hat{\tau}_{hh} } &\approx \EEsub{\mathcal{S}}{ \MSE{\hat{\tau}_{hh} | \mathcal{S}}} + \MSE{ \nu_{\mathcal{S}} } .  \label{eq:MSE_breakdown}
\end{align}

The first term will tend to be a function of the randomization method used and sample-dependent parameters such as $\sigma_S^2(1)$, $\sigma_S^2(0)$, $\sigma_S^2(\Delta)$, and, importantly, the choice of estimator $\hat{\tau}_*$.
For a given choice of $\nu_{\mathcal{S}}$, if we reduce this inner term, we reduce the expectation and therefore increase the overall precision of the estimator for PATE.
We reduce this term with better estimators, e.g., ones that exploit covariates; this is the goal of post-stratification.

The sampling scheme and choice of $\nu_{\mathcal{S}}$ governs the second term.
If we reduce it by changing $\nu_{\mathcal{S}}$, we increase precision.
The main way to do this is to sample better, e.g., move closer to equal probability sampling. 
No estimation strategy can reduce this term.

%
\paragraph{Alternate estimators.} 
Given the above, our primary``double-H\`{a}jek'' estimator $\hat{\tau}_{hh}$ can be viewed as doubly biased: the expected value across randomizations is approximately $\nu_{\mathcal{S}}$, and the expected value of $\nu_{\mathcal{S}}$ is approximately $\tau$.
We could instead use Horvitz-Thompson style estimators at either or both levels to remove these biases.
In particular, if we select an estimator that is unbiased at the randomization level, i.e. $\EE{\tau_* | \mathcal{S} } = \nu_{\mathcal{S}}$, then we have
\begin{align*}
 \MSE{ \hat{\tau}_* } &= \EEsub{\mathcal{S}}{ \var{ \hat{\tau}_* | \mathcal{S} } } + \varsub{\mathcal{S}}{ \nu_{\mathcal{S}} } + \left( \EEsub{\mathcal{S}}{ \nu_{\mathcal{S}} } - \tau \right)^2 
\end{align*}
One such estimator is the ``single-H\`{a}jek'' estimator of
\begin{align*}
 \hat{\tau}_{h} &= \frac{1}{Zp} \sum_{i=1}^N S_i T_i w_i  y_i(1) -  \frac{1}{Z(1-p)}  \sum_{i=1}^N S_i (1-T_i) w_i  y_i(0) .
\end{align*}
This estimator is tied to double-H\`{a}jek by $\EE{Z_1|\mathcal{S}} = pZ$ and $\EE{Z_0|\mathcal{S}} = (1-p)Z$.  It is a Horvitz-Thompson estimator with respect to the randomization for the two parts of our estimand $\nu_{\mathcal{S}}$.
Interestingly, this estimator has the same asymptotic variance expression found in Theorem~\ref{thm:AV_hh} as $\hat{\tau}_{hh}$.

Finally, if $\EEsub{\mathcal{S}}{\nu_{\mathcal{S}}} = \tau$ we have
\begin{align*}
 \MSE{ \hat{\tau}_* } &= \EEsub{\mathcal{S}}{ \var{ \hat{\tau}_* | \mathcal{S} } } + \varsub{\mathcal{S}}{ \nu_{\mathcal{S}} }.
\end{align*}
For fixed $n$, we have such an estimator as
\begin{align}
  \hat{\tau}_{sd} &=  \frac{1}{n_1} \sum_{i \in \mathcal{S}} T_i w_i y_i(1) -  \frac{1}{n - n_1} \sum_{i \in \mathcal{S}} (1-T_i)  w_i y_i(0) \label{eq:tau_sd} .
\end{align}
This estimator generally pays a large price for unbiasedness with high variance.

\section*{Appendix B: Post-Stratification for PATE in Survey Experiments}
\label{sec:post-strat}

Post-stratification is motivated by viewing PATE estimation as a two step process.
In particular, estimators $\nu_{\mathcal{S}}$ that have higher precision will give overall gains.
Say we had a categorical covariate $b$ associated with our outcomes.
We can then express our overall estimand $\tau$ as:
\[ \tau =  \sum_{k=1}^K \frac{N_k}{N}  \left( \frac{1}{N_k} \sum_{i : b_i = k} ( y_i(1) - y_i(0) ) \right) = \sum_{k=1}^K f_k \tau_k \]
with $N_k$ being the number of units in the population in stratum $k$ and $f_k = N_k/N$ being the proportion of the population in stratum $k$.
We could then estimate the population $\tau_k$ with strata level estimators of
\[ \nu_{Sk} = \frac{1}{Z_k} \sum_{i : b_i = k} w_i (y_i(1) - y_i(0)) .\]
As before, we would then need to estimate these $\nu_{Sk}$.

This motivates a post-stratified estimator as a combination of estimates of population strata size estimates and population strata effect estimates:
\[ \hat{\tau}_{ps} = \sum_{k=1}^K \hat{f}_k \hat{\tau}_k  \]
where $\hat{f}_k = Z_k / Z$ estimates $f_k$, with the $Z$ being the total weight in the sample  and the
\[ Z_k = \sum_{i:b_i = k}  w_i S_i  \mbox{ for } k = 1, \ldots, K \]
 being the total weights of the strata. 
These are not dependent on the randomization so 
\[ \EE{ \hat{\tau}_{ps} | \mathcal{S}} = \sum_{k=1}^K \hat{f}_k \EE{\hat{\tau}_k | \mathcal{S}} .\]
If we had population knowledge we might actually know the $f_k$ and simply plug them in; this connects to the generalization of experiments.  
See, for example, \cite{Tipton:2012fj}.

For the $\tau_k$ we have several options.
Arguably the most natural is the double-H\`{a}jek estimator of
\[ \hat{\tau}_k = \frac{1}{Z_{k1}} \sum_{i:b_i = k} S_i T_i w_i y_i(1) - \frac{1}{Z_{k0}} \sum_{i:b_i = k} S_i (1-T_i) w_i y_i(0) \]
with $Z_{k1}$ being the total weight in the treatment group in stratum $k$, and similarly for the control.
The $\hat{\tau}_k$ will have the usual bias from being H\`{a}jek estimators.  
Here, however, this bias is of order $n_k$, not $n$ (see Lemma \ref{lemma:bias_sample}), and so could potentially be larger than one might expect.

Regardless, combining gives our final
\begin{align}
\label{eq:ps}
 \hat{\tau}_{ps}  &=  \sum_{k=1}^K \frac{Z_k}{Z} \left( \frac{1}{Z_{k1}} \sum_{i : b_k = k} S_iT_i w_i y_i(1) - \frac{1}{Z_{k0}} \sum_{i : b_k = k} S_i(1-T_i) w_i y_i(0) \right) .
\end{align}

If we want to avoid this bias, we could instead use a single-H\`{a}jek estimator in each strata:
\begin{align*}
   \hat{\tau}_k^{(h)} & = \frac{n_k}{Z_k} \left( \frac{1}{n_{Tk}} \sum_{i : b_i = k}  T_i w_i  y_i(1) - \frac{1}{n_k - n_{Tk}} \sum_{i : b_i = k}  (1-T_i) w_i  y_i(0)  \right).
\end{align*}
For the single-H\`{a}jek, we immediately have $\EE{ \hat{\tau}_k^{(h)} | \mathcal{S}} = \nu_{S_k}$, i.e., unbiasedness in the randomization step.
This also causes the $Z_k$ to cancel.
If the weights within strata are generally homogenous, the single-H\`{a}jek will be essentially the same as the double. 
And if $b$ is built by stratifying on weights then we would indeed expect such homogeneity.
Thus, with post-stratification, we can remove some bias for very little cost in variance.

%
%

\subsection{Variance Estimation}
\label{sec:bootstrap}

As discussed in the main text, the post-stratification step can be sample-dependent.  
For example, if the units are divided into $K$ quantiles by survey weight, the cut-points of those quantiles depend on the realized weights of the sample.
Because this is still pre-randomization, this does not impact the validity of the variance and variance-estimation formulae of the SATE estimate of $\tau_{\mathcal{S}}$.
It does, however, make generating appropriate population variance formulae difficult.
Furthermore, even if the strata are pre-defined, the formulae of Theorem \ref{thm:AV_hh} are actually for a linearized version of the ratio estimators, and as the strata are smaller than the overall sample, one might be concerned that these approximations would be not that good when applied to individual strata.
This is why we propose the bootstrap. 

Appropriate implementation of the bootstrap deserves some discussion.
Bootstrap is a ``by analogy'' technique.  
To obtain the variability of an estimator we repeatedly simulate obtaining a sample from some population using our hypothesized sampling mechanism, randomizing it into treatment, and estimating the treatment effect using our estimator on that sample.
We first, therefore, need to have a population to sample from.  
Our best estimate of this population is the sample weighted by the weights. 
We then take a size-$n$ i.i.d. sample from this population with probability proportional to the inverse of these weights.
The treatment assignment being Bernoulli means we take a case-wise bootstrap, bootstrapping the original treatment assignment along with the outcome.
This avoids any need to impute any missing potential outcomes.

The up-weighting and subsequent weighted sampling steps collapse to generating a bootstrap sample by taking a classic with-replacement unweighted sample (i.e., a case-wise bootstrap) from the original sample of the triples $( Y^{obs}_i, Z_i, w_i )$.

\section*{Appendix C: Derivations}

In the following we derive the bias of the H\`{a}jek estimator, show that it is small, and derive the bias of $\tau_{SATE}$ as an estimator for the PATE. After this we show how a weighted OLS regression can be used in practice to estimate the double-H\`{a}jek. Finally, we derive properties of the unstratified PATE estimators.

\subsection*{Bias of the H\`{a}jek Estimator}

The proof of Lemma \ref{lemma:bias_sample}, that the bias of a H\`{a}jek estimator is $O(1/\EE{n})$, follows a similar strategy to the proof of Result 6.34 in \cite{Cochran:1977}. That result is of the bias of a general ratio estimator for a fixed sample size under simple random sampling. 
We adapt this result to the H\`{a}jek estimator (also a ratio estimator) under independent Poisson random sampling with variable sample size. A fixed sample size correction is possible, but is not needed for our purposes. 

We extend the notation described in Section \ref{subsec:survey_details}.
Denote $Z_y = \sum_{i = 1}^N \frac{\bar{\pi}}{\pi_i} S_i y_i$ so that we can write $\hat{y}_H = \frac{Z_y}{Z}$. 
The expected values of both the numerator and denominator are
\begin{align*}
	\E[Z_y] &= N\bar{\pi}\mu, \stepcounter{equation}\tag{\theequation}\label{eq:ZyZ}\\
	\EE{Z} &= N\bar{\pi}.
\end{align*}
These results alone should motivate why the H\`{a}jek estimator should be approximately unbiased, but let us be a bit more rigorous.
By first manipulating the difference of the estimator and its target and then applying the first order Taylor approximation, $(1+A)^{-1}\ \dot{=}\ (1-A)$, we can get the approximate difference.
\begin{align*}
	\hat{y}_H - \mu &= \frac{Z_y}{Z} - \mu = \frac{Z_y - \mu Z}{Z} = (Z_y - \mu Z)\frac{1}{Z}\\
	&= (Z_y - \mu Z)\frac{1}{N\bar{\pi}}\frac{N\bar{\pi}}{Z} = (Z_y - \mu Z)\frac{1}{N\bar{\pi}}\left(\frac{Z}{N\bar{\pi}}\right)^{-1}\\
	&= (Z_y - \mu Z)\frac{1}{N\bar{\pi}}\left(\frac{N\bar{\pi} + (Z - N\bar{\pi})}{N\bar{\pi}}\right)^{-1}\\
	&= (Z_y - \mu Z)\frac{1}{N\bar{\pi}}\left(1 + \frac{Z - N\bar{\pi}}{N\bar{\pi}}\right)^{-1}\\
	&\ \dot{=}\ (Z_y - \mu Z)\frac{1}{N\bar{\pi}}\left(1 - \frac{Z - N\bar{\pi}}{N\bar{\pi}}\right)
\end{align*}

Taking expectations and noting that $\E[Z_y - \mu Z] = 0$ by Equation \ref{eq:ZyZ} leads to the approximate bias:
\begin{align*}
	\E[\hat{y}_H] - \mu\ &\dot{=}\ -\frac{1}{(N\bar{\pi})^2}\E\left[(Z_y - \mu Z)(Z - N\bar{\pi})\right] \stepcounter{equation}\tag{\theequation}\label{eq:bias}\\
	&= -\frac{1}{(N\bar{\pi})^2}\left(\E[Z_y Z] - N\bar{\pi}\E[Z_y] + N\bar{\pi} \mu \E[Z] - \mu \E[Z^2]\right).
\end{align*}
These expanded terms can be calculated individually for our estimator using properties of variance and covariance.
\begin{align*}
	\E[Z_y Z] &= Cov(Z_y, Z) + \E[Z_y]\E[Z] \stepcounter{equation}\tag{\theequation}\label{eq:ZZ}\\&= \sum_{i = 1}^N \sum_{j = 1}^N \frac{\bar{\pi}^2}{\pi_i \pi_j}y_i\ Cov(S_i, S_j) + (N\bar{\pi}\mu) (N\bar{\pi})\\
	&=\sum_{i = 1}^N \frac{\bar{\pi}^2}{\pi_i^2}y_i\ Var(S_i) + N^2\bar{\pi}^2\mu \\&= \bar{\pi}^2\sum_{i = 1}^N \frac{1-\pi_i}{\pi_i}y_i + N^2\bar{\pi}^2\mu\\
	&= \bar{\pi}^2\sum_{i = 1}^N \frac{y_i}{\pi_i} - N\bar{\pi}^2\mu + N^2\bar{\pi}^2\mu\\
	\E[Z^2] &= Var(Z) + \E[Z]^2 \stepcounter{equation}\tag{\theequation}\label{eq:Z2}\\&= \sum_{i = 1}^N \frac{\bar{\pi}^2}{\pi_i^2}var(S_i) + N^2\bar{\pi}^2 \\&= \bar{\pi}^2\sum_{i = 1}^N \left(\frac{1}{\pi_i}-1\right) + N^2\bar{\pi}^2 \\
	&= \bar{\pi}^2\sum_{i = 1}^N \frac{1}{\pi_i} - N \bar{\pi}^2 + N^2\bar{\pi}^2 
\end{align*}
Finally, substitute Equations \ref{eq:ZyZ}, \ref{eq:ZZ} and \ref{eq:Z2} into Equation \ref{eq:bias} and simplify:
\begin{align*}
	\E[\hat{y}_H] - \mu\ &\dot{=} -\frac{1}{(N\bar{\pi})^2}\Bigg(\bar{\pi}^2\sum_{i = 1}^N \frac{y_i}{\pi_i} - N\bar{\pi}^2\mu + N^2\bar{\pi}^2\mu \\
	&\ \ \ \ \ \ \ \ \ \ \ \ \ \ \ \ \ \ \ \ \ \ -N^2\bar{\pi}^2\mu + N^2\bar{\pi}^2 \mu \\
	&\ \ \ \ \ \ \ \ \ \ \ \ \ \ \ \ \ \ \ \ \ \ - \mu\bar{\pi}^2\sum_{i = 1}^N \frac{1}{\pi_i} + N \bar{\pi}^2\mu + N^2\bar{\pi}^2 \Bigg)\\
	&= -\frac{1}{N\bar{\pi}} \Bigg( \bar{\pi}\frac{1}{N}\sum_{i = 1}^N \frac{y_i}{\pi_i} - \mu\bar{\pi}\frac{1}{N}\sum_{i = 1}^N \frac{1}{\pi_i} \Bigg)\\
	&= -\frac{1}{\E[n]} \Bigg( \bar{\pi}\frac{1}{N}\sum_{i = 1}^N \frac{y_i - \mu}{\pi_i} \Bigg)\\
	&= -\frac{1}{\E[n]} \Bigg( \frac{1}{N}\sum_{i = 1}^N ({y_i - \mu})\frac{\bar{\pi}}{\pi_i} \Bigg).
\end{align*}
We finally use the relation
\[ \cov{ A, B } = \EE{ (A - \bar{A})(B-\bar{B}) } = \EE{ (A - \bar{A})B } - \EE{ (A - \bar{A}) \bar{B} } = \EE{ (A - \bar{A})B } \]
to get our final covariance formulation.

We have ignored a mild technical issue of an undefined estimator with probability $\pr{ Z = 0 }$.
For the Poisson selection scheme, with the $S_i$ independent, $\pr{ Z = 0 } = \prod (1-\pi_i)$ which will be exponentially small in $n$.
Letting the estimator be defined as 0 under this circumstance gives a bounded, exponentially small term far less in magnitude than other bias terms.

\subsection*{Bias of the SATE for the PATE}

To see that $\hat\tau_{SATE}$ (or $\tau_{\mathcal{S}}$) is a biased estimate for PATE, assume fixed sample size $n$ to obtain:
\begin{align*}
 \EE{  \hat{\tau}_{SATE} } = \EEsub{\mathcal{S}}{  \EE{  \hat{\tau}_{SATE} | \mathcal{S} } } = \EEsub{\mathcal{S}}{ \tau_{\mathcal{S}} }  &= \EEsub{\mathcal{S}}{ \frac{1}{n} \sum_{i=1}^N S_i (y_i(1)-y_i(0)) } \\
 &= \frac{1}{N} \sum_{i=1}^N \frac{N \pi_i}{n} \left( y_i(1) - y_i(0) \right).
\end{align*}
For a random sample size, there is an additional, but negligible, a bias term. 
We can see that the above is a first order approximation of the overall bias by replacing $\EEsub{\mathcal{S}}{ \rfrac{S_i}{n}}$ with $\rfrac{\EEsub{\mathcal{S}}{S_i}}{\EEsub{\mathcal{S}}{n}}$.
The difference in these terms is of order $1/n$, as with our bias lemma.

\subsection*{The double-H\`{a}jek as weighted OLS} 
\label{sub:double_hajek_as_weighted_ols}

In Section \ref{sec:estimating_nuS} we introduced the ``double-H\`{a}jek'' estimator. Here we show that this estimate is equivalent to a weighted OLS where the weights are $w_i = \frac{\bar{\pi}}{\pi_i}$ and we regress on the treatment indicator. In other words we fit the model
\begin{align*}
	y_i = \alpha + \tau T_i + \varepsilon_i
\end{align*}
with weights $w_i$. 
The weighted OLS estimates $\hat\alpha$ and $\hat\tau$ are the solutions to the normal equations:
\begin{align}
	\sum_{i\in S} w_i(y_i - \hat\alpha - \hat\tau T_i) = 0,\\
	\sum_{i\in S} w_i T_i (y_i - \hat\alpha - \hat\tau T_i) = 0. \label{eq:normal_2}
\end{align}
These are obtained by taking derivatives with respect to $\alpha$ and $\tau$ of the weighted sum of squares, $\sum_{i\in S} w_i(y_i - \alpha - \tau T_i)^2$, and setting them to $0$. Grouping by treatment indicators, we get the following:
\begin{align*}
	& \sum_{i: T_i = 1} w_i (y_i - \hat\alpha - \hat\tau) + \sum_{i: T_i = 0} w_i (y_i - \hat\alpha) = 0,\\
	& \sum_{i: T_i = 1} w_i (y_i - \hat\alpha - \hat\tau) = 0.
\end{align*}
Taking the difference of these equations implies that
\begin{align*}
	\hat\alpha = \frac{\sum_{i: T_i = 0} w_i y_i}{\sum_{i: T_i = 0} w_i}.
\end{align*}
To make the connection to the ``double-H\`{a}jek'' estimate, denote $Z_0 = \sum_{i: T_i = 0} w_i$ and $Z_1 = \sum_{i: T_i = 1} w_i$, as before. 
If we distribute the summation in the second normal equation (Equation~\ref{eq:normal_2}), we get
\begin{align*}
		& \sum_{i: T_i = 1} w_i y_i - \hat\alpha Z_1  - \hat\tau Z_1  = 0\\
		& \sum_{i: T_i = 1} w_i y_i - \frac{Z_1}{Z_0} \sum_{i: T_i = 0} w_i y_i  - \hat\tau Z_1  = 0\\
		& \hat\tau = \frac{1}{Z_1}\sum_{i: T_i = 1} w_i y_i - \frac{1}{Z_0} \sum_{i: T_i = 0} w_i y_i \\
\end{align*}
Written in the most general sense and replacing the weights, we get back our ``double-H\`{a}jek'' estimate.
\begin{align*}
	\hat\tau_{hh} &=  \frac{1}{Z_1} \sum_{i=1}^N S_i T_i \frac{\bar{\pi}}{\pi_i}  y_i(1) -  \frac{1}{Z_0}  \sum_{i=1}^N S_i (1-T_i) \frac{\bar{\pi}}{\pi_i}  y_i(0)
\end{align*}

Hence one way of calculating $\hat{\tau}_{hh}$ is by fitting a weighted OLS regression onto the treatment indicator and inspecting the coefficients. 


\subsection*{Properties of \texorpdfstring{$\hat\tau_{hh}$}{tau hh} }

Our estimator can be expressed as
\begin{align*}
 \hat{\tau}_{hh} &=  \frac{1}{Z_1} \sum_{i=1}^N S_i T_i \frac{\bar{\pi}}{\pi_i} y_i(1) -  \frac{1}{Z_0}  \sum_{i=1}^N S_i (1-T_i) \frac{\bar{\pi}}{\pi_i} y_i(0) \\
   &= \hat{\mu}(1) - \hat{\mu}(0) .
\end{align*}
For the expectation of $\hat{\tau}_{hh}$, we have
\begin{align*}
 \EE{ \hat{\tau}_{hh} | \mathcal{S}}  &\approx  \EE{ \frac{1}{\EE{Z_1|\mathcal{S}}} \sum_{i=1}^N S_i  T_i \frac{\bar{\pi}}{\pi_i}  y_i(1) -  \frac{1}{\EE{Z_0|\mathcal{S}}}  \sum_{i=1}^N S_i (1-T_i) \frac{\bar{\pi}}{\pi_i}  y_i(0) | \mathcal{S} } \\
 &=  \frac{1}{Z} \sum_{i=1}^N S_i  \frac{\bar{\pi}}{\pi_i}  y_i(1) -  \frac{1}{Z}  \sum_{i=1}^N S_i  \frac{\bar{\pi}}{\pi_i}  y_i(0) = \nu_{\mathcal{S}} 
\end{align*}

For variance we use results and notation from \cite{sarndal2003model} to obtain approximate variance terms as follows.
Define $\tilde{S}_i = S_i T_i$ as the event of unit $i$ being selected and also treated.
We then have $\tilde{\pi}_i = \EE{ \tilde{S}_i } = p \pi_i$ and the probability that units $j$ and $k$ are both selected and treated is
\[ \tilde{\pi}_{jk} = \EE{ \tilde{S}_j = 1 \mbox{ and } \tilde{S}_k = 1 } = \pr{ T_j = 1 \mbox{ and } T_k = 1 | S_j = 1, S_k = 1} \pi_{jk}  \]

For the treatment group specifically we have
\[  \hat{\mu}(1)  =  \frac{\bar{\pi} \sum_{i=1}^N S_i T_i \frac{y_i(1)}{p \pi_i} }{ \bar{\pi} \sum_{i=1}^N S_i T_i \frac{1}{p \pi_i}  } 
		 = \frac{\sum_{i=1}^N S_i T_i \frac{y_i(1)}{p \pi_i} }{ \sum_{i=1}^N S_i T_i \frac{r_i}{p \pi_i}  } 
		 	= \frac{ \sum_{\tilde{S}} \check{y}_i }{ \sum_{\tilde{S}} \check{r}_i } = \frac{ \hat{t}_y }{ \hat{t}_r }.
			 \]
with $r_i = 1$. The check notation denotes a value divided by its probability of being included in the sample: $\check{a}_i = a_i / \pi_i$.
 The above is a classic ratio estimator with selection probabilities of $\tilde{\pi}_j$ for the ratio of
\[ R = \frac{ t_y }{ t_r } =  \frac{ \sum_{i=1}^N y_i(1) }{ \sum_{i=1}^N r_i } = \frac{\sum_{i=1}^N y_i(1)}{N}  = \mu(1) \]
since $t_r = \sum_{i=1}^N r_i = N$.

The approximate variance of a ratio estimator \citep{sarndal2003model} is:
\begin{align*}
AV(  \hat{\mu}(1) ) &= \frac{1}{t^2_r} \sum_{j=1}^N \sum_{k=1}^N \tilde{\Delta}_{jk} \frac{ y_j(1) - R r_j }{ \tilde{\pi}_j } \frac{ y_k(1) - R r_k }{ \tilde{\pi}_k } \\
 	&= \frac{1}{N^2} \sum \sum \tilde{\Delta}_{jk} \frac{ y_j(1) - \mu(1)}{ p\pi_j }  \frac{ y_k(1) -  \mu(1) }{ p\pi_k } \\
 	&= \frac{1}{N^2} \sum \sum \frac{\tilde{\Delta}_{jk}}{p^2 \pi_j\pi_k} \left(y_j(1) - \mu(1)\right)\left(y_k(1) -  \mu(1)\right) \\
\end{align*}
with
\[ \tilde{\Delta}_{jk} \equiv \tilde{\pi}_{jk} - \tilde{\pi}_j \tilde{\pi}_k = \tilde{\pi}_{jk} - p^2 \pi_j \pi_k .\]

We can estimate this variance with a sum over the treatment group of
\begin{align*}
\widehat{V}(  \hat{\mu}(1) ) &= \frac{1}{\widehat{N}^2} \sum_{j=1}^N \sum_{k=1}^N S_j T_j S_k T_k\frac{\tilde{\Delta}_{jk}}{\tilde{\pi}_{jk}p^2 \pi_j\pi_k} \left(y_j(1) - \hat\mu(1)\right)\left(y_k(1) -  \hat\mu(1)\right)\\
\end{align*}
with $\hat\mu(1) = \frac{1}{\widehat{N}}\sum_{i = 1}^N S_i T_i \check{y}_i(1)$ and $\widehat{N} = \sum S_i T_i / \pi_i p$.

\paragraph{The Poisson-Bernoulli Model.}
Under Poisson selection  we have $\pi_{jk} = \pi_j \pi_k$ for $j \neq k$ (with $\pi_{jj} = \pi_j$).  With Bernoulli assignment we have $\tilde{\pi}_{jk} = p^2 \pi_j \pi_k$ for $j \neq k$ (with $\tilde{\pi}_{jj} = p \pi_j$) giving $\tilde{\Delta}_{jk} = 0$ for $j \neq k$ and $\tilde{\Delta}_{jj} = p \pi_j(1- p\pi_j)$ for $j=k$.  This gives
\begin{align*}
AV(  \hat{\mu}(1) ) &= \frac{1}{N^2} \sum_{j=1}^N  \frac{1 - p\pi_j}{p\pi_j}\left(y_j(1) - \mu(1) \right)^2 
\end{align*}
and
\begin{align*}
\widehat{V}(  \hat{\mu}(1) ) &= \frac{1}{\widehat{N}^2}  \sum_{j = 1}^N S_j T_j \frac{ 1 - p\pi_j }{ p^2\pi_j^2} \left(y_j(1) -  \hat{\mu}(1) \right)^2 .
\end{align*}

The above formula are problematic in that they depend on our $\pi_j$ rather than the weights $w_j = \bar{pi}/\pi_j$).  
However, if we assume $N \gg n$ we can make progress.
In particular, in this case, under mild regularity conditions on the sampling probabilities, we can assume $\pi_j \ll 1$ for all $j$. 
This means that $1 - p\pi_j \approx 1$.
Couple this with $N \bar{\pi} = \EE{n}$ to get a fairly tight upper bound on our two formula of
\begin{align*}
AV(  \hat{\mu}(1) ) &\leq \frac{1}{p \EE{n}} \frac{1}{N} \sum_{j=1}^N  w_j \left(y_j(1) - \mu(1) \right)^2 
\end{align*}
and, using $\widehat{N} = Z_1 / (\bar{\pi} p)$ with $Z_1 = \sum S_j T_j w_j$,
\begin{align*}
\widehat{V}(  \hat{\mu}(1) ) &= \frac{ \bar{\pi}^2}{Z_1^2}  \sum_{j = 1}^N S_j T_j \frac{ 1 - p\pi_j }{\pi_j^2} \left(y_j(1) -  \hat{\mu}(1) \right)^2 \\
 &\leq \frac{1}{Z_1^2} \sum_{j = 1}^N S_j T_j w_j^2 \left(y_j(1) -  \hat{\mu}(1) \right)^2 .
\end{align*}

Finally, to get overall variance presented in Theorem~\ref{thm:AV_hh} we first view the sample into the treatment arm as independent of the sample into the control arm, which is again motivated by the $N \gg n$ assumption.
For the control arm, we then do the above derivation with $\tilde{S}_i = S_i(1-T_i)$ and $\tilde{\pi}_i = (1-p)\pi_i$.
More lengthy derivations that account for the dependence structure will give higher-order terms which are in the end negligible.
See \citet{wood2008covariance} for an approach.

%

\section*{Appendix D: The simulation's DGP}

\begin{figure}[ht!]
\centering
\begin{subfigure}{\textwidth}
  \centering
  \includegraphics[scale=.52,page=1]{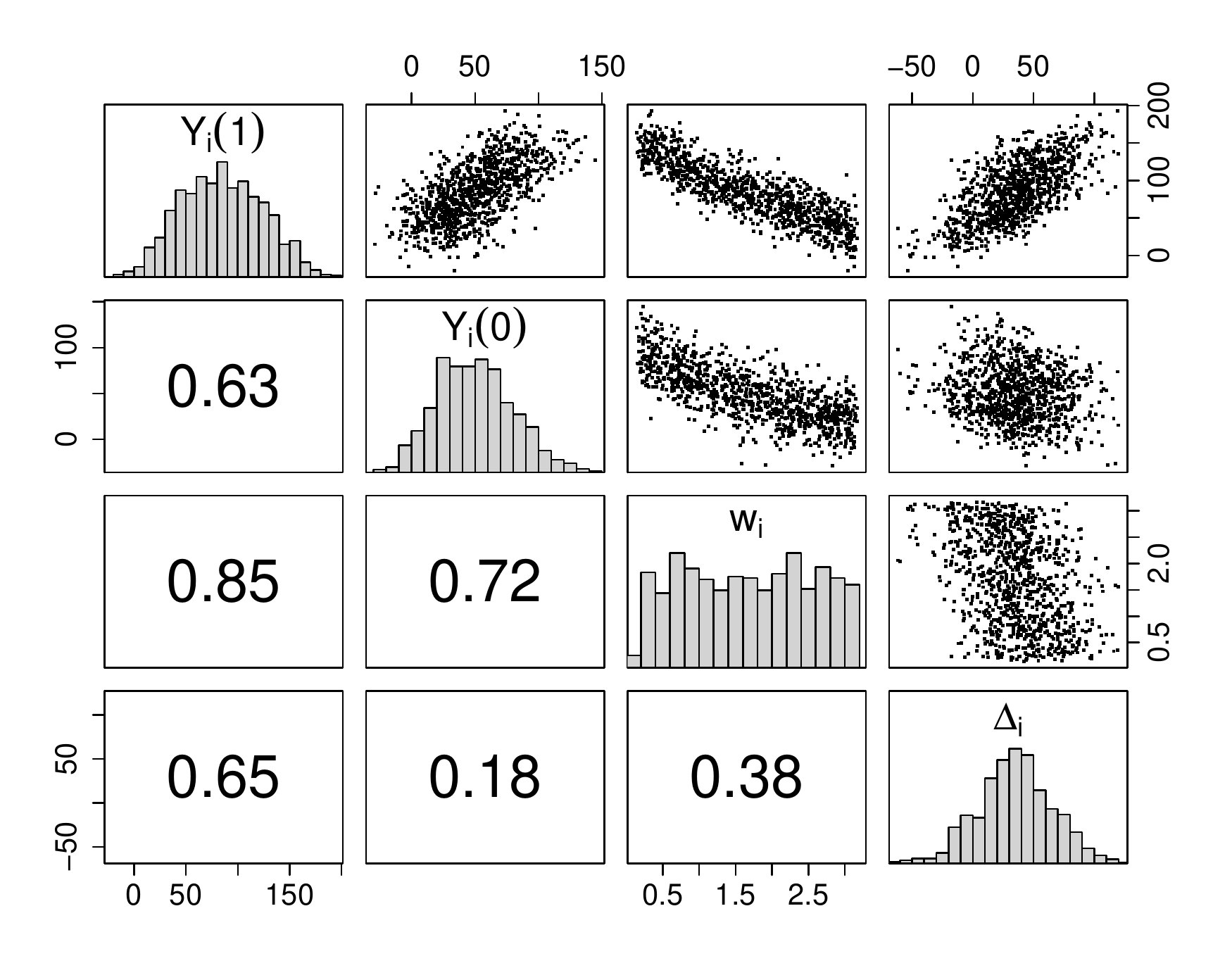}
  \caption{Population characteristics for Simulation A where the heterogeneous treatment effect varies in connection to the weight. $Y_i(1)$ and $Y_i(0)$ are respectively the treatment and control potential outcomes, $w_i$ is the unit weight (units are sampled inversely proportional to this) and $\Delta_i$ is the individual treatment effect.}
  \label{fig:simpopulation_pop}
\end{subfigure}\\
\begin{subfigure}{\textwidth}
  \centering
  \includegraphics[scale=.52,page=2]{figuresSimulationA.pdf}
  \caption{Characteristics of a sample from Simulation A. $b_i$ is the post-stratification generated on this particular sample.
}
  \label{fig:simpopulation_sample}
\end{subfigure}

\caption{Characteristics of the Population and a Sample from Simulation A}
\label{fig:simpopulation}
\end{figure}

In this section we provide additional simulation details and explanations of some of the choices made throughout the simulations of Section \ref{sec:simulations}. 
In all our simulations, the potential outcomes are simulated as nonlinear functions of the weights. 

To generate our populations we use the following algorithm: let $\gamma \in [0,1]$ be a correlation measuring the strength of the relationship between the weights and outcomes.
We then generate two latent parameters $(\varepsilon_{i},\tilde{\varepsilon}_{i})$ as a bivariate standard normal draw with correlation $\gamma$.
(We do this by generating $\varepsilon_{i} \sim N( 0, 1)$, and $\tilde{\varepsilon} = \gamma \varepsilon_{i} + \sqrt{ 1 - \gamma^2 } \eta_i$, with $\eta_i \sim N(0,1)$.)

We then generate uniformly distributed weights on pre-specified interval $(a, b)$ by using the c.d.f. transformation:
\[ w_i = a + b\ \Phi(\varepsilon_{i}), \]
where $\Phi$ is the standard normal c.d.f.
We also generate shadow weights
\[ \tilde{w}_i = a + b\ \Phi(\tilde{\varepsilon}_{i}) ,\]
also uniform, and with the same distribution as $w_i$.

Our potential outcomes are then a function of the shadow weights $\tilde{w}_i$:
\begin{align*}
 Y_i(0) &= 120 - 20 \sqrt{\tilde{w}_i} + 5 \epsilon_i \\
 Y_i(1) &= Y_i(0) + 10\sqrt{b - \tilde{w}_i} 
\end{align*}
with $\epsilon_i$ as independent Gaussian noise.
The treatment potential outcomes are generated to give a non-linear heterogeneous treatment effect. 
When $\gamma = 1$, $\tilde w_i = w_i$, giving the strongest possible relationship between outcome and weight.
Conversely when $\gamma = 0$ the weights are completely unrelated to the potential outcomes, so stratifying on them should not help improve estimation.

Once we have a population, we then sample inversely proportional to the weight $w_i$.
For example, in Simulation A we take a fixed sample size of $n = 500$ ($5\%$ of the population). 
Our post-stratification estimator stratifies based on the weight $w_i$ to increase precision. 
The stratifying variable $b_i$ is defined in Section \ref{sec:post-strat}.

Simulation A has maximal covariance, with $\gamma=1$.
Figure \ref{fig:simpopulation} shows a subset of the population and a sample from this scenario to illustrate the structure of our DGP.
Figure \ref{fig:simpopulation_pop} shows the characteristics of the simulated population while Figure \ref{fig:simpopulation_sample} shows how a weighted sample might look. 

Overall, Figure \ref{fig:simpopulation} shows that the weight $w_i$ and potential outcome distributions differ in the sample and population. Furthermore, because the potential outcomes are related to the weights they are consequently related to the post-stratification levels $b_i$ in the sample.

For Simulation B we simply replace the formula for $Y_i(1)$ with a constant treatment effect of $30$, so $Y_i(1) = Y_i(0) + 30$. 
We still have the sample general relationships between the sample and population, but as we see in Section \ref{sec:simulations} the estimators behave quite differently. 

For Simulation C we varied $\gamma$, which controls the relationship between the weight and the potential outcomes.
The top two right-most panels of Figure~\ref{fig:simpopulation_sample} show there is smaller variability within strata for $Y_i(0)$ and $Y_i(1)$ than if we consider the entire sample at once. 
As our weights become less predictive of outcome, this variability will increase.
Our formulation, however, maintains the marginal distributions of $w_i$, $Y_i(0)$, and $Y_i(1)$ as $\gamma$ changes so that any benefits we see from post-stratification can only be attributed to the changing relationship.

\section*{Appendix E: Further Details and Results of the Real Data Application}

As mentioned in the main text, the 92 survey experiments analyzed in Section \ref{sec:application} were generated from 18 unique randomizations on 7 separate surveys. 
We split each randomization by subject party identification and considered multiple outcomes per treatment randomization. 
One might worry that the potential correlation of the multiple outcomes might be influencing the results, so we append here the results when considering only one unique outcome per randomization. 

\begin{figure}[ht!]
\centering
\includegraphics[scale=.52]{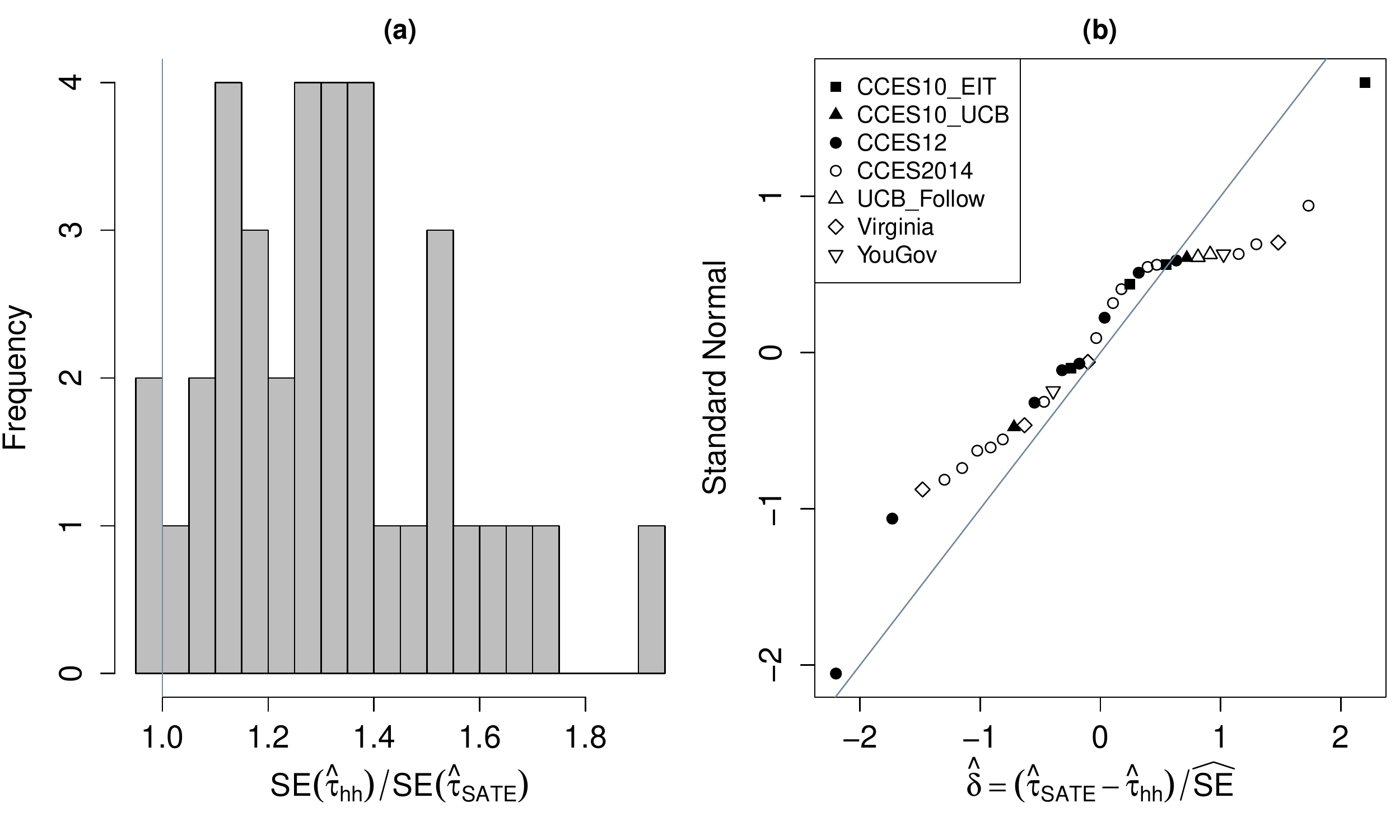}
\caption{(a) Standardized efficiency of estimates of $\hat\tau_{hh}$ vs $\hat\tau_{SATE}$. (b) quantile-quantile comparison plot of the {\emph{relative difference}} $\hat{\delta}$ of the estimates for the 36 experiments grouped by containing survey.}
\label{fig:application_plot_appendix}
\end{figure}

The 18 unique randomizations give rise to 36 survey experiments after splitting each randomization by subject party identification, considering only the larger Democratic and Republican leaning subgroups. 
$28$ of them ($77.8\%$) showed SATEs that were significantly different from zero. 
Once the weights were taken into account to estimate the PATE (via the double-H\`{a}jek estimate) $25$ experiments ($69.4\%$) had significant effects. 
Even though more experiments showed significant PATE than SATE estimates, incorporating weights still increased standard errors: there was a $31.7\%$ average increase in variance of $\hat\tau_{hh}$ over $\hat\tau_{SATE}$ across experiments.
The raw SE increases can be seen in Figure \ref{fig:application_plot_appendix}(a). 

We further examined whether there is evidence of some experiments having a PATE substantially different from the SATE. 
We calculated the 36 $\hat\delta$ values and compared them to a standard normal with a qq-plot (Figure \ref{fig:application_plot_appendix}(b)). 
While visually there do seem to be some distributional departures from a standard Normal, a Kolmogorov–Smirnov test does not support this hypothesis (with a p-value of $0.14$). 
Furthermore, an FDR test also fails to find any experiments with significant differences. 
All of this suggests a general equivalence between the SATE and the PATE in this subset of experiments as well. 

To explore whether post-stratification on weights improved precision, we compared the estimated SEs. The estimated SEs of $\hat\tau_{ps}$ are very similar to those for $\hat\tau_{hh}$, with an average increase of about $0.3\%$. 
Post-stratifying on party ID on the original 18 experiments led to modest variance reduction.
Relative to no stratification, we see an average reduction of $1.6\%$ in variance across experiments with participants of both major parties.
If we post-stratify on both party ID and the weights, we see an average reduction of $1.4\%$. 
These findings, similar to the main text, show that while post-stratification should help reduce the variance in theory the gains can be rather modest in practice.

\begin{sidewaystable}[]
\centering
\caption{Details for Studies Used}
\label{studydetails}
\resizebox{\textwidth}{!}{%
\begin{tabular}{llllllll}
\multicolumn{1}{c}{\cellcolor[HTML]{EFEFEF}Survey} & \multicolumn{1}{c}{\cellcolor[HTML]{EFEFEF}Year} & \multicolumn{1}{c}{\cellcolor[HTML]{EFEFEF}Geography} & \multicolumn{1}{c}{\cellcolor[HTML]{EFEFEF}Base Stimulus} & \multicolumn{1}{c}{\cellcolor[HTML]{EFEFEF}Conditions} & \multicolumn{1}{c}{\cellcolor[HTML]{EFEFEF}Outcome(s)}                                                                                            & \multicolumn{1}{c}{\cellcolor[HTML]{EFEFEF}N Democrats} & \multicolumn{1}{c}{\cellcolor[HTML]{EFEFEF}N Republicans} \\
CCES Module 1                                      & 2010                                             & National                                              & News Report                                               & Democratic/Republican Candidate                        & Report Fair; Report Biased; Topic Important; Candidate Deserves Credit; Candidate Typical                                                                                                 & 231                                                     & 211                                                       \\
CCES Module 1                                      & 2010                                             & National                                              & Image                                                     & Democratic Donkey/Republican Elephant                  & Unemployment Rate Estimate                                                                                                & 190                                                     & 219                                                       \\
CCES Module 2                                      & 2010                                             & National                                              & Image                                                     & Democratic Donkey/Republican Elephant                  & Unemployment Rate Estimate                                                                                                & 201                                                     & 193                                                       \\
YouGov Study                                       & 2011                                             & National                                              & News Report                                               & Democratic/Republican Candidate                        & Report Fair; Report Biased; Topic Important; Candidate Deserves Credit; Candidate Typical                                                                                                 & 442                                                     & 372                                                       \\
CCES                                               & 2012                                             & National                                              & Campaign Advertisement Video                              & Obama/Romney                                           & Time watched; Repeat; Share; See More                                                                                     & 326                                                     & 228                                                       \\
CCES                                               & 2012                                             & National                                              & Campaign Advertisement Video                              & Negative/Positive                                      & Time watched; Repeat; Share; See More                                                                                     & 326                                                     & 228                                                       \\
CCES                                               & 2012                                             & National                                              & Candidate Vignette                                        & Democratic/Republican Label                            & Trait Ratings: Compassionate; Moral; Strong Leader; Really Cares; Knowledgeable; Greedy; Indecisive; Hard Working; Honest & 321                                                     & 225                                                       \\
CCES                                               & 2012                                             & National                                              & Voter Fraud Hypothetical                                  & Democrats/Republicans                                  & Would this group commit fraud?                                                                                            & 223                                                     & 145                                                       \\
Gubernatorial Election                             & 2013                                             & Virginia                                              & Campaign Advertisement Video                              & McCauliffe/Cuccinelli                                  & Time watched; Repeat; Share; See More                                                                                                                            & 454                                                     & 350                                                       \\
Gubernatorial Election                             & 2013                                             & Virginia                                              & Campaign Advertisement Video                              & Negative/Positive                                      & Time watched; Repeat; Share; See More                                                                                                                            & 454                                                     & 350                                                       \\
YouGov Study                                       & 2013                                             & National                                              & News Report                                               & Democratic/Republican Candidate                        & Report Fair; Report Biased; Topic Important; Candidate Deserves Credit; Candidate Typical                                 & 456                                                     & 353                                                       \\
CCES                                               & 2014                                             & National                                              & Candidate Conjoint 1          & Male/Female Candidate                                  & Is candidate more likely a Democrat or Republican?                                                                        & 504                                                     & 330                                                       \\
CCES                                               & 2014                                             & National                                              & Candidate Conjoint 2          & Male/Female Candidate                                  & Is candidate more likely a Democrat or Republican?                                                                        & 504                                                     & 330                                                       \\
CCES                                               & 2014                                             & National                                              & Candidate Conjoint 3          & Male/Female Candidate                                  & Is candidate more likely a Democrat or Republican?                                                                        & 504                                                     & 330                                                       \\
CCES                                               & 2014                                             & National                                              & Candidate Conjoint 4          & Male/Female Candidate                                  & Is candidate more likely a Democrat or Republican?                                                                        & 504                                                     & 330                                                       \\
CCES                                               & 2014                                             & National                                              & Painting by George W. Bush                                & Bush Revealed as Artist/Not Revealed                   & Rating of Painting Quality                                                                                                & 504                                                     & 329                                                       \\
CCES                                               & 2014                                             & National                                              & Sketch by Barack Obama                                    & Obama Revealed as Artist/Not Revealed                  & Rating of Sketch Quality                                                                                                  & 394                                                     & 278                                                       \\
CCES                                               & 2014                                             & National                                              & News Story about Stampede at July 4 Gathering             & Democratic/Republican Event                            & In-Party Shame                                                                                                            & 338                                                     & 204                                                       \\
                                                   &                                                  &                                                       &                                                           &                                                        &                                                                                                                           &                                                         &                                                           \\
                                                   &                                                  &                                                       &                                                           &                                                        &                                                                                                                           &                                                         &                                                           \\
                                                   &                                                  &                                                       &                                                           &                                                        &                                                                                                                           &                                                         &                                                           \\
                                                   &                                                  &                                                       &                                                           &                                                        &                                                                                                                           &                                                         &                                                           \\
                                                   &                                                  &                                                       &                                                           &                                                        &                                                                                                                           &                                                         &                                                          
\end{tabular}%
}
\end{sidewaystable}

\end{document}